\documentclass{article}
\usepackage{spconf,amsmath,amssymb,graphicx}

\usepackage{booktabs}
\usepackage{bm}
\usepackage{xcolor}
\usepackage[hidelinks]{hyperref}
\hypersetup{colorlinks=false}

\title{
Deep network series for large-scale high-dynamic range imaging
}
\name{Amir Aghabiglou$^{\star}$, Matthieu Terris$^{\star}$, Adrian Jackson$^{\ddagger}$, Yves Wiaux$^{\star}$\thanks{The work of A.~A., M.~T., and Y.~W.~was supported by EPSRC under grants EP/T028270/1 and ST/W000970/1. Computing \mbox{resources} came from the Cirrus UK National Tier-2 HPC Service at EPCC (http://www.cirrus.ac.uk) funded by the University of Edinburgh and EPSRC (EP/P020267/1), partly through time allocation under the SUSA project, and partly through GPU resources directly provided by EPCC. A.~J.~was supported by EPSRC under grant EP/T028351/1. A.~A.~and M.~T.~contributed equally.}}
\address{\small$^{\star}$ Institute of Sensors, Signals and Systems, Heriot-Watt University, Edinburgh EH14 4AS, United Kingdom \\
\small$^{\ddagger}$ EPCC, University of Edinburgh, Edinburgh EH8 9BT, United Kingdom}
\begin{document}
%
\maketitle
\begin{abstract}
We propose a new approach for large-scale high-dynamic range computational imaging. Deep Neural Networks (DNNs) trained end-to-end can solve linear inverse imaging problems almost instantaneously. While unfolded architectures provide robustness to measurement setting variations, embedding large-scale measurement operators in DNN architectures is impractical. Alternative Plug-and-Play (PnP) approaches, where the denoising DNNs are blind to the measurement setting, have proven effective to address scalability and high-dynamic range challenges, but rely on highly iterative algorithms.
We propose a residual DNN series approach, also interpretable as a learned version of matching pursuit, where the reconstructed image is a sum of residual images progressively increasing the dynamic range, and estimated iteratively by DNNs taking the back-projected data residual of the previous iteration as input. 
We demonstrate on radio-astronomical imaging simulations that a series of only few terms provides a reconstruction quality competitive with PnP, at a fraction of the cost.
\end{abstract}

\section{Introduction}
\label{sec:intro}

Designing algorithms for solving ill-posed imaging inverse problems that both deliver high precision reconstructions and scale to large data volumes is a significant challenge of imaging sciences.
A problem where such considerations are paramount is aperture synthesis by Radio Interferometry (RI) in astronomy, an iconic example of linear inverse Fourier imaging problems \cite{wiaux2009compressed}.
In this context, the acquisition of an unknown target radio image $\bm{\bar{x}}\in\mathbb{R}^N$ from observed data $\bm{y}\in\mathbb{C}^M$ follows the observation model
\begin{equation}
\bm{y}=\bm{\mathsf{\Phi}} \bm{\bar{x}}+\bm{n},
\label{eq:inv_pb}
\end{equation}
where $\bm{n}\sim\mathcal{N}(0,\tau^2)$ is the realisation of a complex Gaussian random noise, and where the linear measurement operator $\bm{\mathsf{\Phi}}\colon \mathbb{R}^N \to \mathbb{C}^M$ represents (in its simplest form, see \cite{terris2022image} for details) a non-uniform Fourier transform. 
More precisely, $\bm{\mathsf{\Phi}} = \bm{\mathsf{UFZ}}$, where $\bm{\mathsf{Z}}$ is a zero-padding operator, $\bm{\mathsf{F}}$ denotes the 2D discrete Fourier transform, and $\bm{\mathsf{U}}$ is an interpolation matrix mapping the discrete Fourier coefficients onto the measured values of the continuous Fourier transform. The set of sampled values is characterised by a so-called sampling pattern. With the advent of a whole new generation of radio telescopes arrays aiming to probe the sky with much higher resolution and sensitivity, such as the flagship MeerKAT (\href{https://www.sarao.ac.za/}{sarao.ac.za}) and SKA (\href{https://www.skatelescope.org/}{skatelescope.org}), extreme image dimensions and data volumes are to be expected, along with dynamic ranges (\emph{i.e.}~the ratio between the brightest and faintest sources in $\bm{\bar{x}}$) spanning multiple orders of magnitude. 

Deep Neural Networks (DNNs) have shown outstanding results in solving linear inverse imaging problems. On the one hand, end-to-end approaches provide extremely fast reconstruction. They are widespread in other imaging communities \cite{ahmad2020plug, muckley2021results, liang2021swinir}, but their use remains limited in RI imaging \cite{terris2022image, gheller2021convolutional, connor2022deep}. This is mainly due to the lack of ground-truth datasets, combined with a wide intrinsic variability of the RI observation model, leading to generalisation issues. While unfolded architectures \cite{adler2018learned, banert2020data} provide necessary robustness to variations of the measurement setting, embedding large-scale measurement operators in DNN architectures is impractical, both for training and inference. On the other hand, Plug-and-Play (PnP) algorithms \cite{venkatakrishnan2013plug, zhang2021plug, pesquet2021learning}, substituting learned DNN denoisers in lieu of proximal regularisation operators in optimisation algorithms, have shown outstanding performance and robustness, including for high-dynamic range imaging \cite{terris2022image,dabbech2022first}. However, PnP approaches remain highly iterative and will still struggle to scale to the image sizes and data volumes of interest in applications such as RI imaging.

In this paper, a new residual DNN series approach to large-scale high-dynamic range computational imaging is devised, where the reconstructed image is built as a sum of few
residual images progressively increasing the dynamic range, and estimated iteratively as output of DNNs taking the back-projected data residual of the previous iteration as input. We show on preliminary simulations for RI imaging that such a strategy may enable to achieve an imaging quality matching that delivered by the most advanced optimisation and PnP algorithms, while  being orders of magnitude faster thanks to its very limited iterative nature.

\section{Methodology}
\label{sec:approach}

\subsection{Proposed approach}
\label{subsec:proposedapproach}

We propose to train a series of DNNs $(\bm{\mathsf{G}}^{(i)})_{1\leq i \leq I}$ yielding a sequence of estimated reconstructions $(\bm{x}^{(i)})_{1\leq i \leq I}$ of $\bm{\bar{x}}$, following the update rule

\begin{equation}
(\forall 1\leq i \leq I) \quad
\left\{
\begin{aligned} 
\bm{r}^{(i-1)}&=\bm{x}_{\textrm{dirty}}-\kappa \textnormal{Re}\{\bm {\mathsf{\Phi}}^{\dagger}{\bm{\mathsf{\Phi}}}\}\bm{x}^{(i-1)},\\ 
\bm{x}^{(i)}  &= \bm{x}^{(i-1)} +  \bm{\mathsf{G}}^{(i)} \left( \bm{r}^{(i-1)} \right),
\end{aligned}
\right.
\label{eq:sum_rules}
\end{equation}
where $\bm{x}^{(0)}=0$ and where $(\bm{r}^{i-1})_{1\leq i \leq I}$ corresponds to the data residuals back-projected in the image domain. The back-projected or so-called dirty image reads $\bm{x}_{\textrm{dirty}} = \kappa \textnormal{Re}\{\bm{\mathsf{\Phi}}^\dagger\bm{y}\}$.

\begin{figure}[t]
\includegraphics[width=.48\textwidth, 
clip]{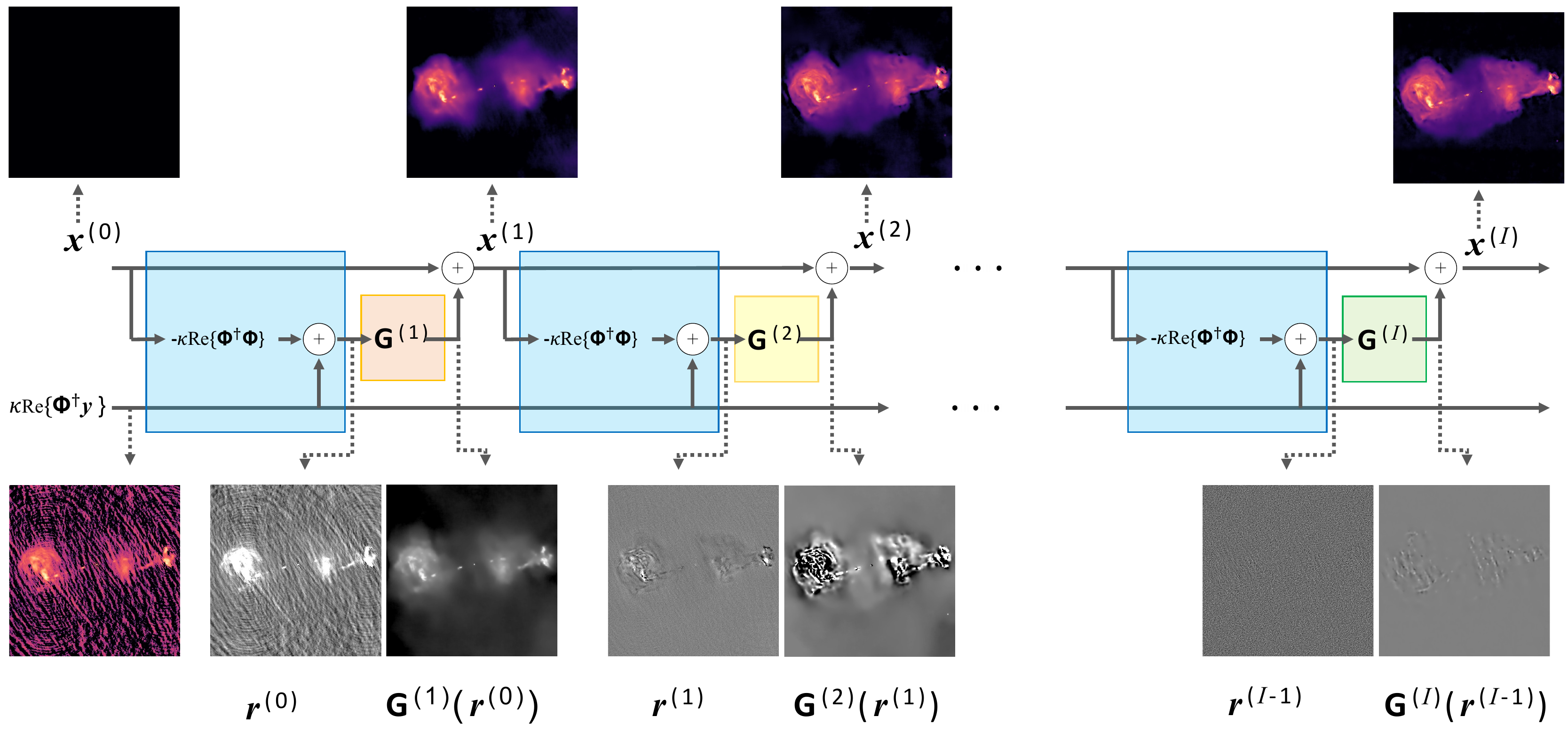}
\vspace{-2.25em}
\caption{\small Summary of the proposed approach. Top row shows the successive image estimates $\bm{x}^{(i)}$ for $1\leq i \leq I$ while the bottom row shows the back-projected data residuals $\bm{r}^{(i-1)}$, input to $\bm{\mathsf{G}}^{(i)}$, and the predicted image residuals $\bm{x}^{(i)}-\bm{x}^{(i-1)}$, output of $\bm{\mathsf{G}}^{(i)}$. Each network is shown in a different colour to stress that it is trained separately. In blue, the computation of $\bm{r}^{(i-1)}$. Note that $\bm{x}^{(0)}=0$ and $\bm{r}^{(0)}=\kappa\textnormal{Re}\{\bm{\mathsf{\Phi}}^\dagger\bm{y}\}$.}
\vspace{-1.25em}
\label{fig:res}
\end{figure}

In \eqref{eq:sum_rules}, 
we set $\kappa = 1/\max(\textnormal{Re}\{\bm{\mathsf{\Phi}}^{\dagger}\bm{\mathsf{\Phi}}
\}\bm{\mathsf{\delta}})$ to ensure the normalisation of back-projected data residuals at the input of the networks ($\delta$ being the Dirac delta). 
The last element of the sequence 
and final reconstruction
$\bm{x}^{(i)}$ also reads
\begin{equation}
{\bm x}^{(I)}= {\sum_{i=1}^{I}}\bm{\mathsf{G}}^{(i)}(\bm{r}^{(i-1)}),
\label{eq:sum_interp}
\end{equation}
hence the name ``deep network series''. In practice, we aim at keeping $I$ as small as possible. We also note that the first iteration is equivalent to a standard end-to-end learning approach, as $\bm{x}^{(1)}=\bm{\mathsf{G}}^{(1)}(\bm{r}^{(0)})$ with $\bm{r}^{(0)}=\bm{x}_{\textrm{dirty}}$ the back-projected data. The approach is summarised in Figure~\ref{fig:res}.

In order to train the sequence of DNNs $(\bm{\mathsf{G}}^{(i)})_{1\leq i \leq I}$, we start from a dataset of $L$ synthetic ground-truth images $(\bm{\bar{x}}_{l})_{1\leq l \leq L}$ from which we simulate measurements $(\bm{y}_{l})_{1\leq l \leq L}$ as per \eqref{eq:inv_pb}. Then, for every $1\leq i \leq I$, we generate a dataset of residuals $(\bm{r}_{l}^{(i-1)})_{1\leq l\leq L}$ as per \eqref{eq:sum_rules} and train $\bm{\mathsf{G}}^{(i)}$ to
\begin{equation}
\underset{\bm{\theta}}{\min}\dfrac{1}{L} \sum_{l=1}^{L} \lVert \bm{\mathsf{G}}^{(i)}_{\bm{\theta}}(\bm{r}_{l}^{(i-1)})+\bm{x}_{l}^{(i-1)}-\bm{\bar{x}}_{l} \rVert_{1},
\end{equation}
where $\bm{\theta}$ denotes the learnable parameters of $\bm{\mathsf{G}}^{(i)}$. 

\subsection{Relation to unfolding, PnP, and matching pursuit}

The proposed approach is reminiscent of both unfolded and PnP algorithms based on the Forward-Backward optimisation algorithm \cite{bauschke2017convex}, as the back-projected data residuals in \eqref{eq:sum_rules} can be written as $\bm{r}^{(i-1)}=-\kappa\nabla f(\bm{x}^{(i-1)})$ for $f(\cdot) = \frac{1}{2}\|\bm{\mathsf{\Phi}}\cdot-\bm{y}\|_2$. There are however important differences with such schemes. Firstly, instead of applying $\bm{\mathsf{G}}^{(i)}$ to $\operatorname{Id}-\kappa \nabla f$, $\bm{\mathsf{G}}^{(i)}$ is applied to the residual $-\kappa \nabla f$ only. Secondly, unfolded architectures are trained in an end-to-end fashion, whereas our approach proposes to train $\bm{\mathsf{G}}^{(i)}$ sequentially, with a dataset of back-projected data residuals $(\bm{r}_{l}^{(i-1)})_{1\leq l\leq L}$ as input. We stress that this procedure, involving the training of $I$ DNNs, is dictated by the large-scale nature of $\bm{\mathsf{\Phi}}$, precluding its direct embedding into a DNN architecture. Thirdly, DNNs in PnP algorithms are generally the same across iterations and trained as simple denoisers. 

In a nutshell, our approach can be interpreted as a learned version of matching pursuit, iteratively identifying model components from back-projected data residuals. While \cite{Hauptmann8327873} adopts a similar approach, their simple network architecture explicitly requires the current estimate as second input, and the approach is not investigated at high dynamic range.

\section{Experiments}
\label{sec:experiments}

\subsection{Training RI dataset \& training}
\label{dataset}

To circumvent the absence of an appropriate dataset of target radio images, we rely on a synthetic dataset from \cite{terris2022image}, where noisy, low-dynamic range optical images have been processed through denoising and exponentiation procedures to create a dataset of $L=2235$ clean images $(\bm{\bar{x}}_{l})_{1\leq l \leq L}$, of size $N = 512\times 512$, and with nominal dynamic range slightly above $10^4$. The full dataset is normalised so that the maximum pixel value across all images is equal to 1.

For each image $\bm{\bar{x}}_l$ we create data $\bm{y}_l = \bm{\mathsf{\Phi}}_l\bm{\bar{x}}_l+\bm{n}_l$, with a measurement operator $\bm{\mathsf{\Phi}}_l$ resulting from simulating a sampling pattern of the MeerKAT telescope, with fixed integration and observation times, and randomly chosen pointing direction, leading to data vectors of size $M = 1.6\times 10^6$. The measurement noise $\bm{n}_l$ is generated with fixed standard deviation set to ensure a back-projected noise level $\tau/ \sqrt{2\|\textnormal{Re}\{{\bm{\mathsf{\Phi}}_l}^\dagger \bm{\mathsf{\Phi}}_l\}\|_2}\lesssim 10^{-4}$, thus leading to a target dynamic range of the order of, or slightly above, $10^4$. Following standard practice in the field, the data $\bm{y}_l$, operator $\bm{\mathsf{\Phi}}_l$, and noise $\bm{n}_l$ are weighted component-wise by the inverse square-root of the sampling density at each point to mitigate the side-lobes of the point spread function $\textnormal{Re}\{{\bm{\mathsf{\Phi}}_l}^\dagger \bm{\mathsf{\Phi}}_l\}\bm{\delta}$ (see Figure~\ref{fig:step_visualisation}.c).

We train the sequence of networks $(\bm{\mathsf{G}}^{(i)})_{1\leq i \leq I}$ following the procedure detailed in Section~\ref{subsec:proposedapproach}. Each network $\bm{\mathsf{G}}^{(i)}$ has the same architecture, namely a UNet \cite{zhang2021plug} in which convolutional blocks were replaced by WDSR blocks \cite{yu2018wide}. The networks also include an initial normalisation layer, ensuring that any input image effectively has zero mean and unit standard deviation, mitigating potential generalisation issues related to variations of statistics between the training samples and any test sample. The output image is de-normalised as part of a last layer, using the same mean and standard deviation as computed at the input. Each network is trained with the RMSprop algorithm, for approximately 200 epochs for each network. In practice, we noticed that the reconstruction quality saturates for $I>4$ on the training dataset, and therefore chose $I=4$ for all our experiments.

\subsection{Test RI dataset \& simulation set up}

We validate the proposed methodology on a test dataset of 3 radio images (3c353, Hercules A, and Centaurus A, see \cite{terris2022image} for more details) of size $N = 512\times 512$, with peak value normalised to 1, and dynamic range slightly above $10^4$. For each of them, 5 RI data vectors are simulated from 5 MeerKAT sampling patterns generated as in Section \ref{dataset}, using the same integration and observation time, same noise standard deviation, and random pointing directions. This leads to a total of 15 inverse problems with data vectors of size $M = 1.6\times 10^6$. 

Reconstruction quality is evaluated with the signal-to-noise ratio metric, as $\text{SNR}(\bm{x},\bm{\bar{x}}) = 20\operatorname{log}_{10}(\|\bm{\bar{x}}\|_2/\|\bm{\bar{x}}-\bm{x}\|_2)$ for an estimate $\bm{x}$ of $\bm{\bar{x}}$.
Since we aim at reconstructing high-dynamic range images, we visualise reconstructions in logarithmic scale (\emph{i.e.}~after applying the transform $\operatorname{rlog}\colon \bm{x} \mapsto \operatorname{log}_{10}(10^3\bm{x}+1)/3$), which helps visualising very faint features. We also compute the SNR in logarithmic scale as $\text{logSNR}(\bm{x},\bm{\bar{x}}) = \text{SNR}(\operatorname{rlog}(\bm{x}),\operatorname{rlog}(\bm{\bar{x}}))$. When evaluating the reconstruction quality of the proposed method across iterations, we also use the relative norm of the back-projected data residuals $\eta^{(i)}=\|\bm{r}^{(i)}\|_2/\|\bm{r}^{(0)}\|_2$, measuring the relative amount of data left to be processed after each iteration. 

\subsection{Benchmark RI algorithms}

We compare the proposed method to various RI imaging algorithms, from the recent and advanced sparsity-promoting optimisation algorithm SARA \cite{onose2016scalable, thouvenin2022parallel}, to its PnP counterpart AIRI \cite{terris2022image, dabbech2022first}, to the traditional RI imaging algorithm CLEAN \cite{offringa2014wsclean} (relying on a classical matching pursuit approach), and to an end-to-end approach resulting from cutting the network series to $I=1$ in the proposed approach.

\subsection{Results}


\begin{figure}[t] \small %
\centering %
\begin{tabular}{@{\hspace{-0.\tabcolsep}} c @{\hspace{0.1\tabcolsep}} c @{\hspace{0.1\tabcolsep}} c @{\hspace{0.1\tabcolsep}} c @{\hspace{0.\tabcolsep}} l @{\hspace{0.\tabcolsep}}}
\includegraphics[width=0.11\textwidth]{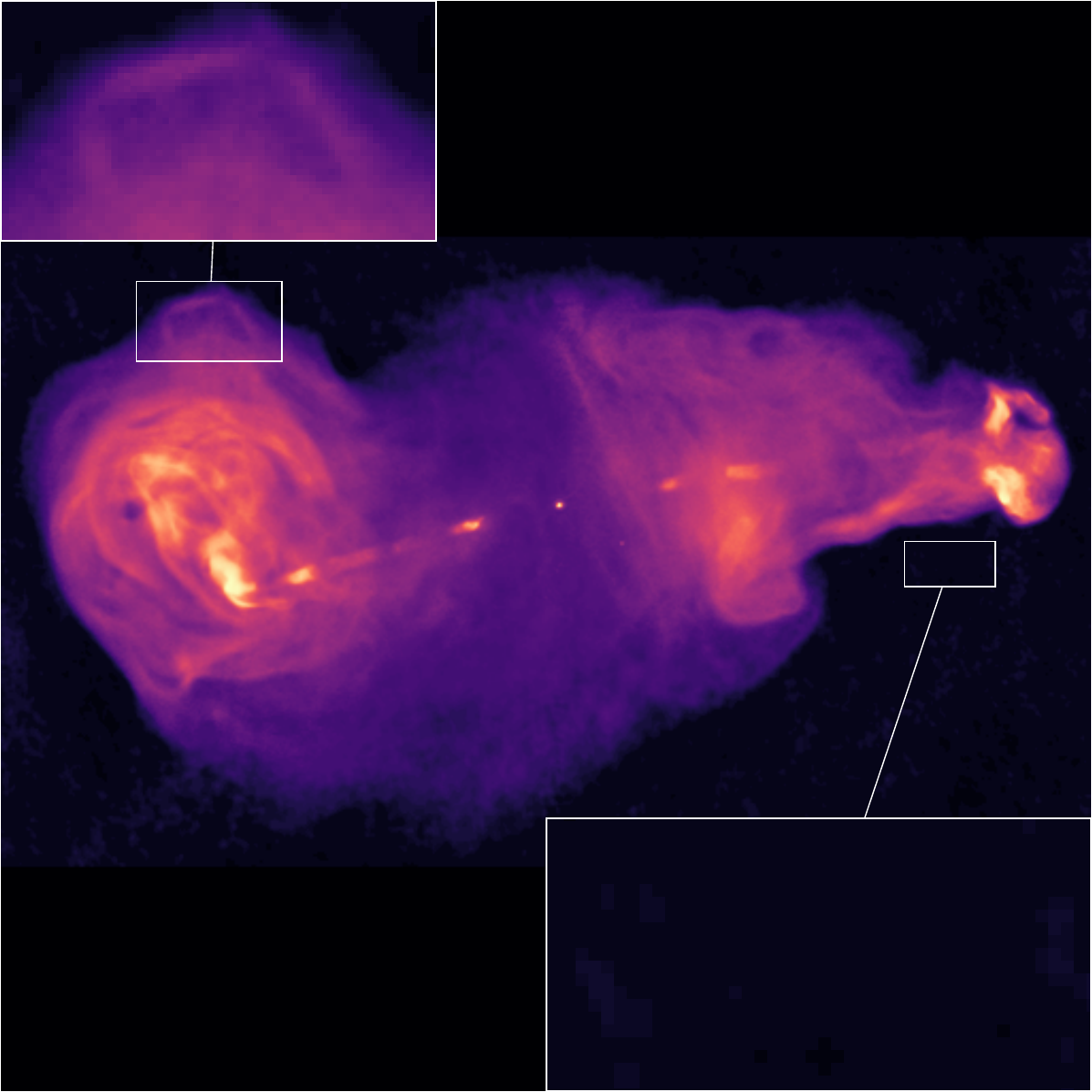} &
\includegraphics[width=0.11\textwidth]{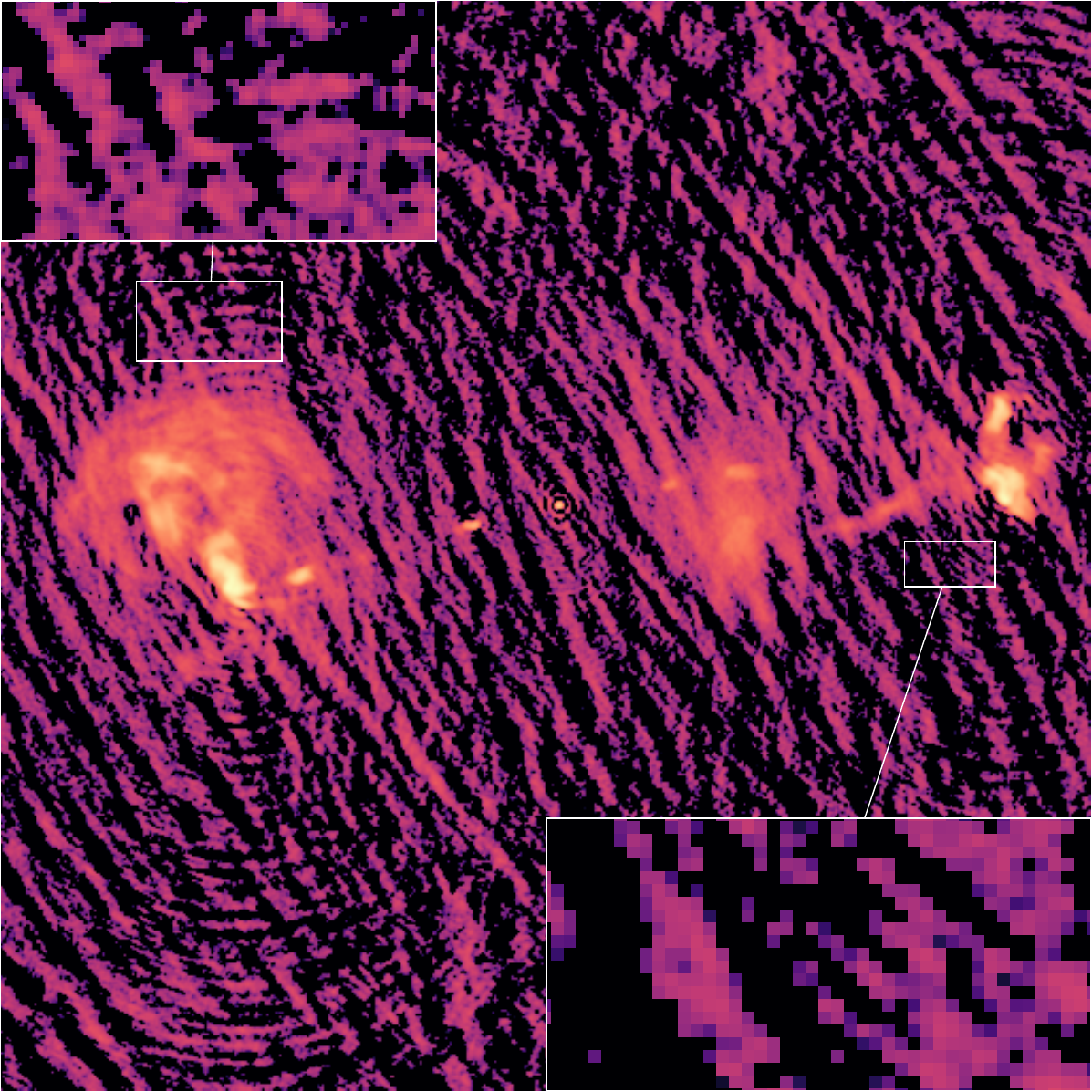} &
\includegraphics[height=0.11\textwidth]{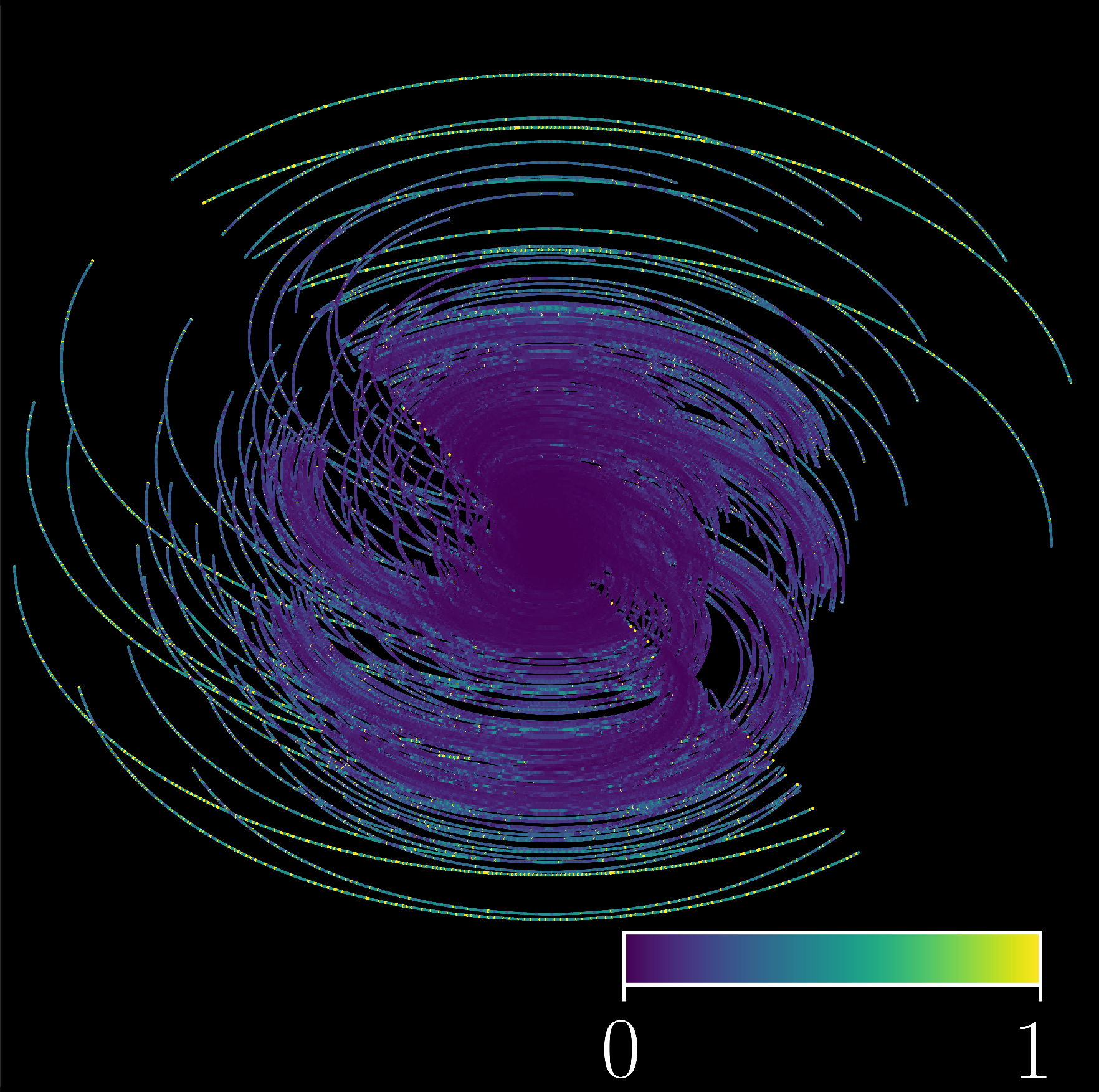}
&
 &
\\
(a) $\bm{\bar{x}}$ & (b) $\kappa\textnormal{Re}\{\bm{\mathsf{\Phi}}^\dagger\bm{y}\}$ & (c) sampling & & \\
\includegraphics[width=0.11\textwidth]{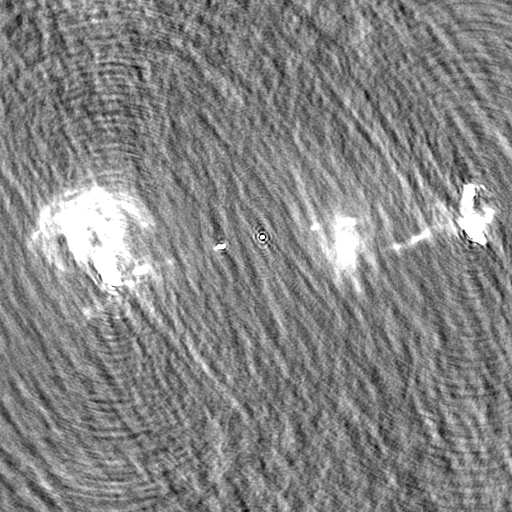} &
\includegraphics[width=0.11\textwidth]{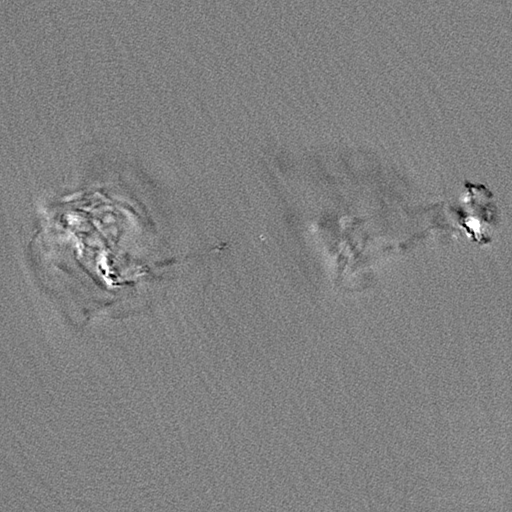} &
\includegraphics[width=0.11\textwidth]{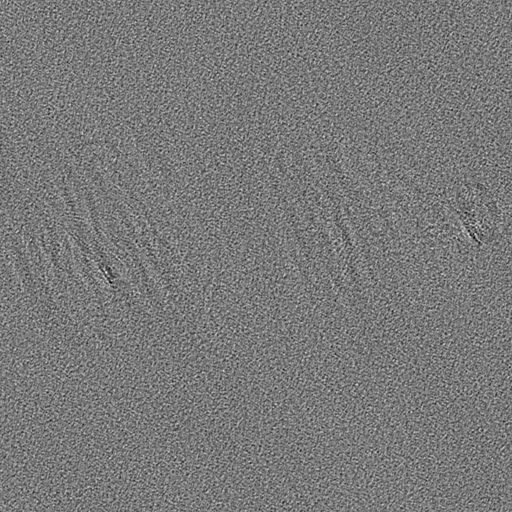} &
\includegraphics[width=0.11\textwidth]{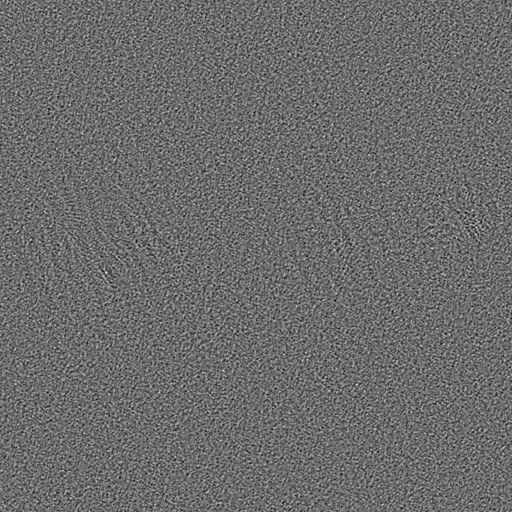} &
\raisebox{-0.\height}[0pt][0pt]{\includegraphics[width=0.025\textwidth]{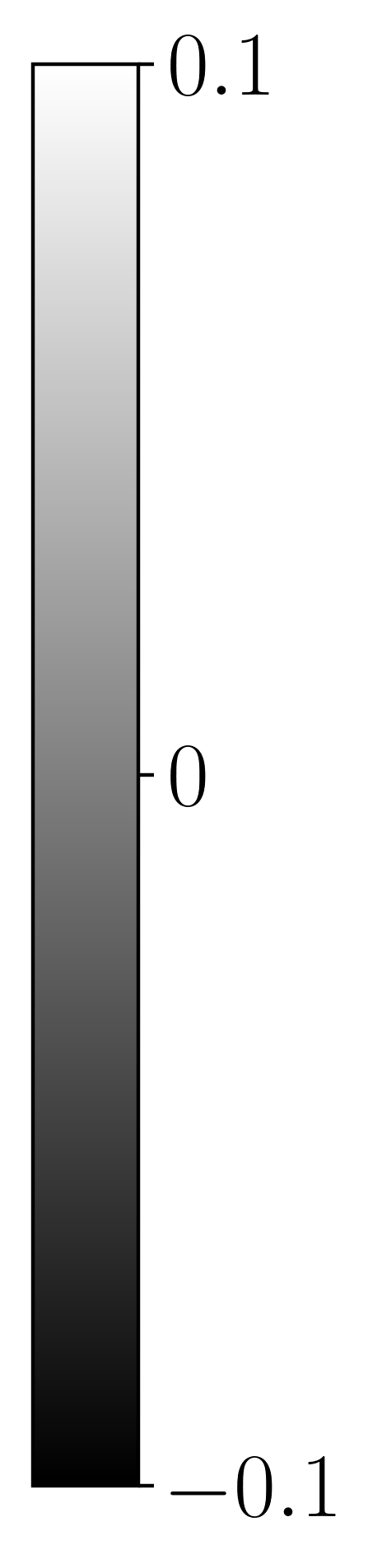}} \\
(d) $\bm{r}^{(0)}=\bm{x}_{\textrm{dirty}}$ & (e) $\bm{r}^{(1)}$ & (f) $\bm{r}^{(2)}$ & (g) $\bm{r}^{(3)}$ \\
\includegraphics[width=0.11\textwidth]{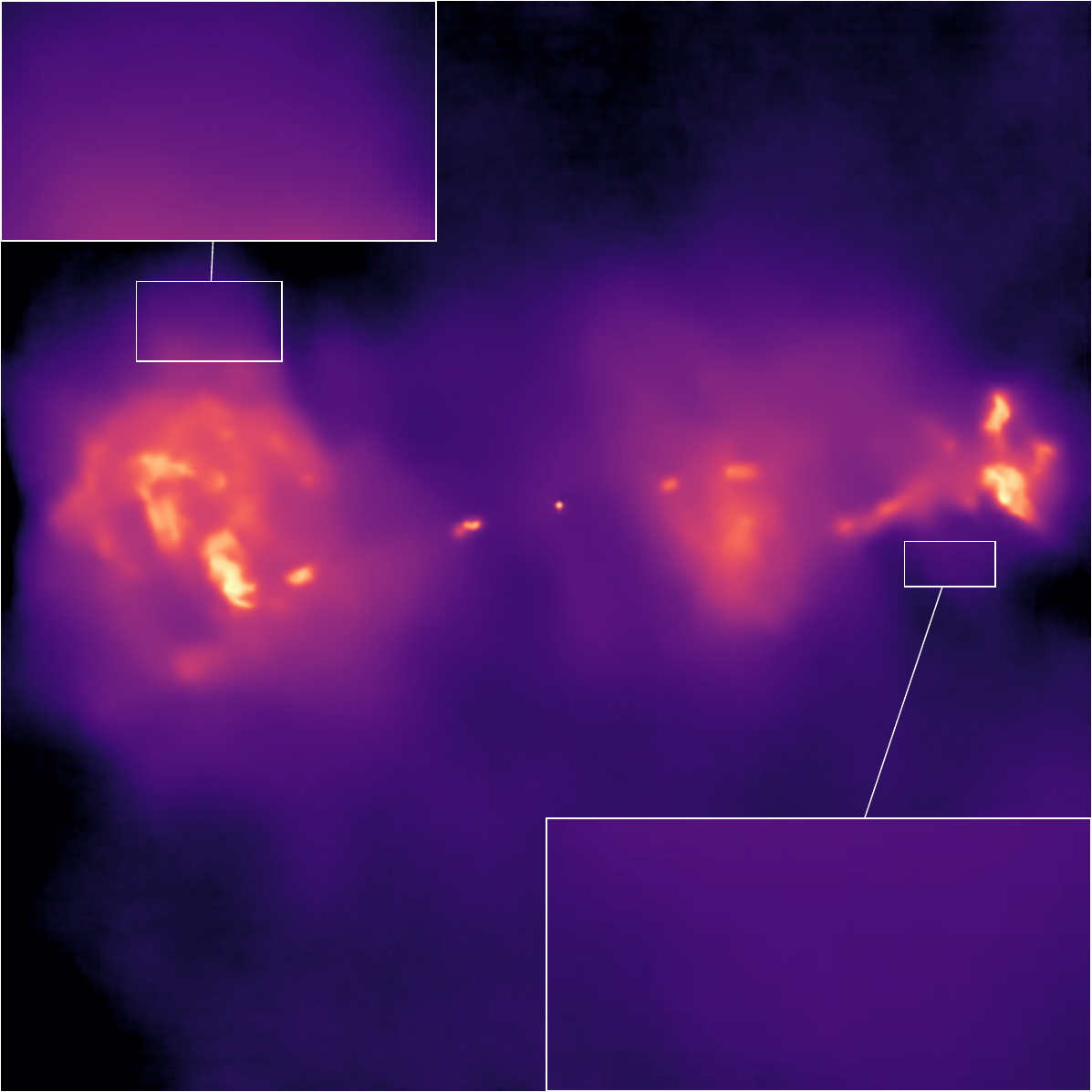} &
\includegraphics[width=0.11\textwidth]{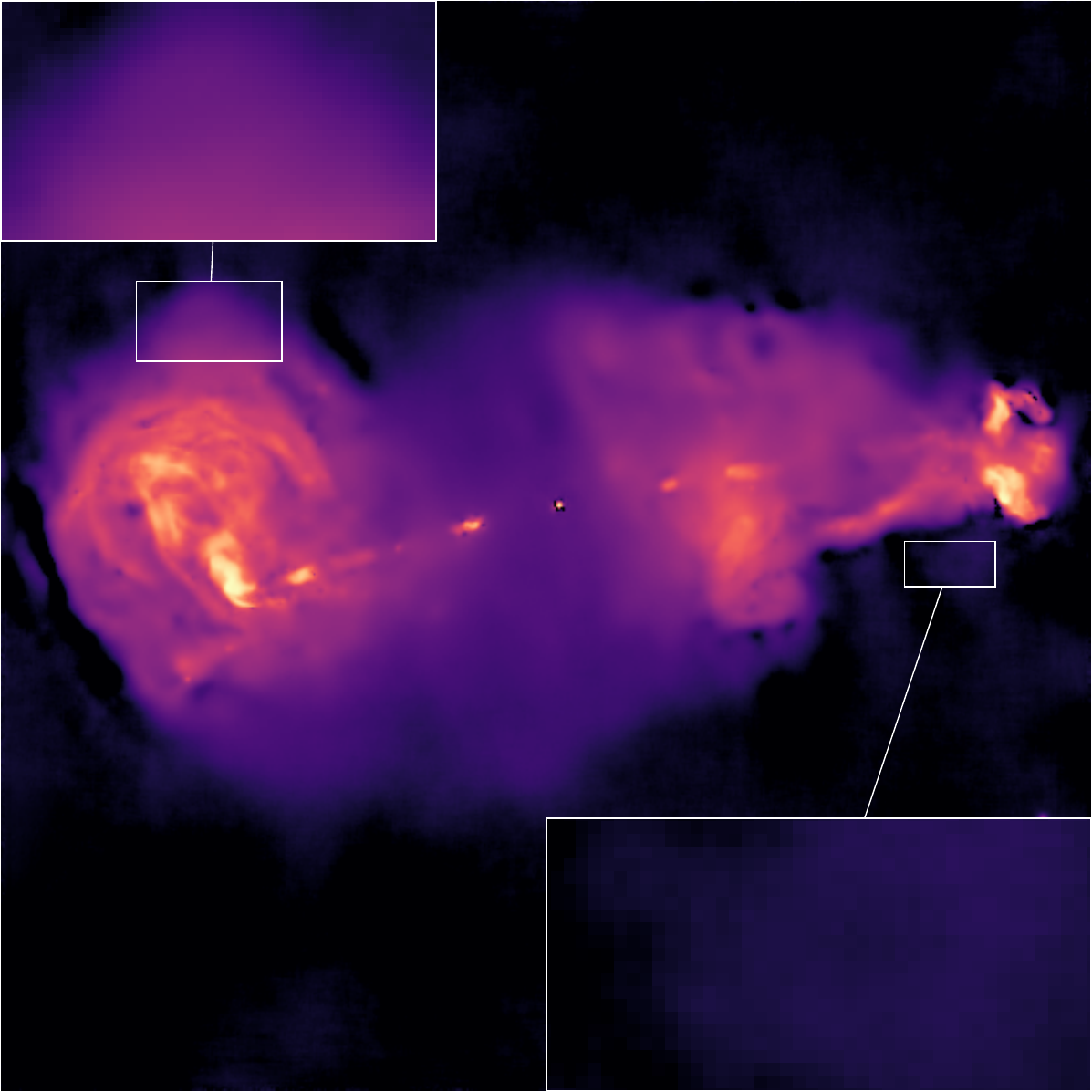} &
\includegraphics[width=0.11\textwidth]{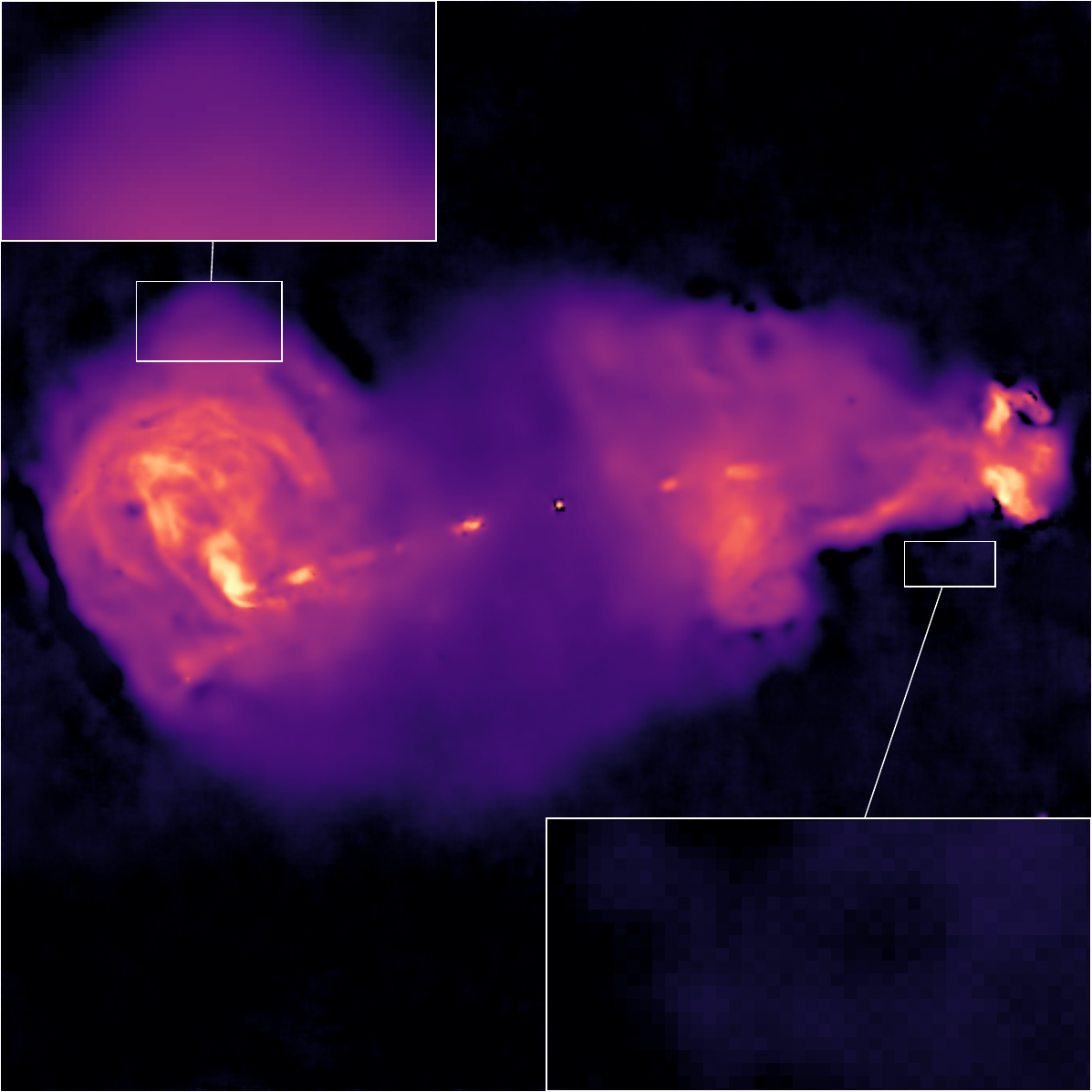} &
\includegraphics[width=0.11\textwidth]{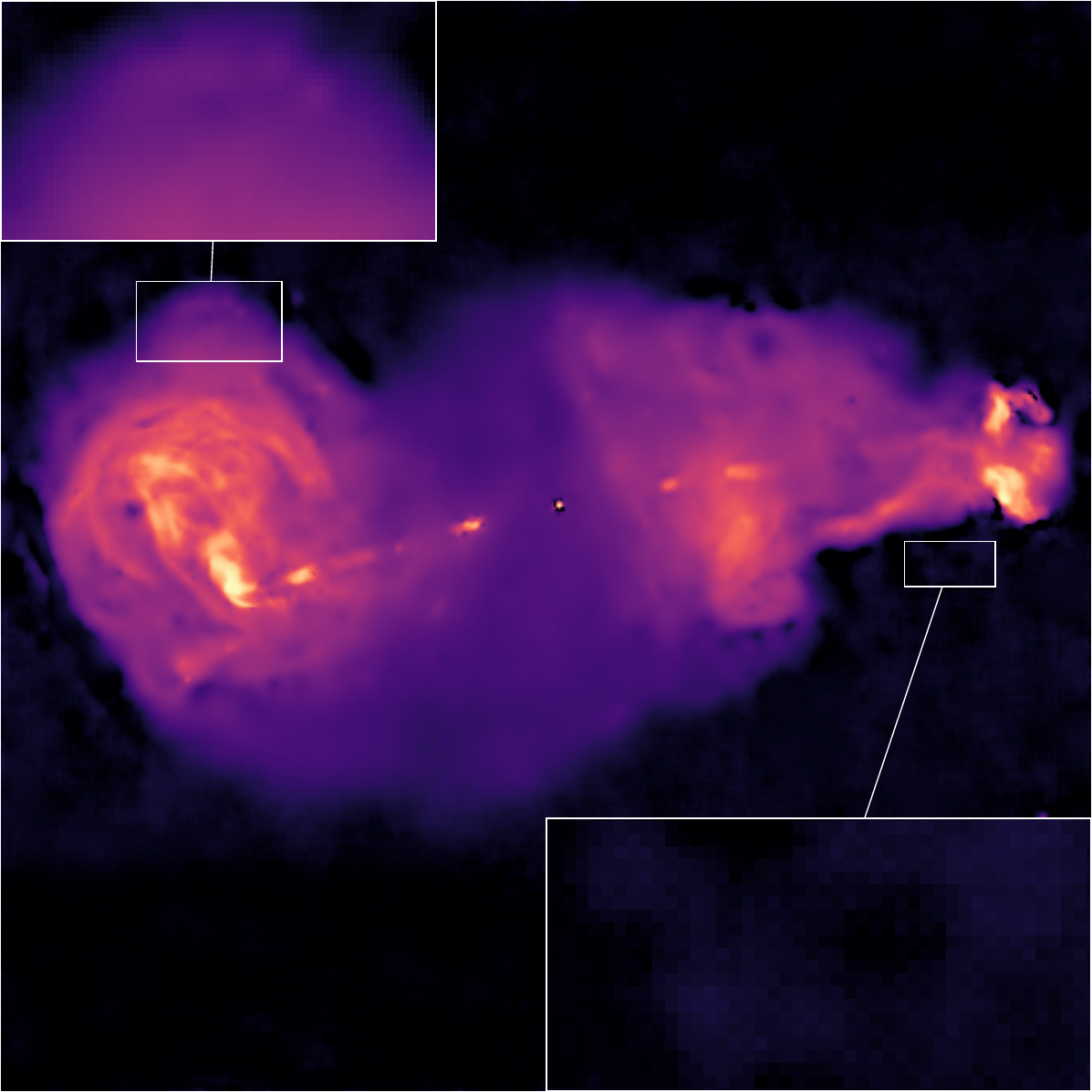} &
\raisebox{-0.\height}[0pt][0pt]{\includegraphics[width=0.025\textwidth]{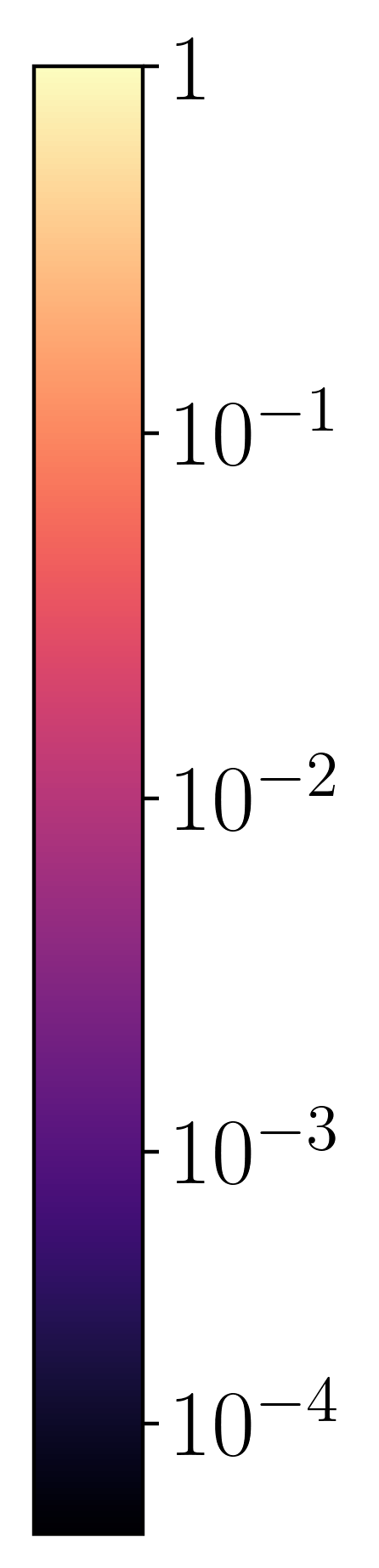}} \\
(h) $\bm{x}^{(1)}$ & (i) $\bm{x}^{(2)}$ & (j) $\bm{x}^{(3)}$ & (k) $\bm{x}^{(4)}$ & \\
$(21.0, 11.0)$ & $(27.9, 22.3)$ & $(28.1, 24.2)$ &
$(28.2, 25.1)$
\end{tabular}
\vspace{-0.5em}
\caption{\small Different stages of the proposed approach illustrated on one of the three images (3c353) and for one of the five sampling patterns of the test dataset. Top row shows (a) the ground truth $\bm{\bar{x}}$, (b) the back-projected data $\kappa\textnormal{Re}\{\bm{\mathsf{\Phi}}^\dagger\bm{y}\}$, and (c) the weighted Fourier sampling pattern (colorbar displaying weight values). The second row (panels (d)-(g)) and third row (panels (h)-(k)) show the back-projected data residuals $\bm{r}^{(i-1)}$ and reconstructions $\bm{x}^{(i)}$ for $1\leq i \leq 4$, below which associated (SNR, logSNR) metrics are displayed.}
\label{fig:step_visualisation}
\end{figure}


\begin{figure}[t] %
\begin{tabular}{@{\hspace{-0.\tabcolsep}} c @{\hspace{-0.\tabcolsep}} c @{\hspace{-0.\tabcolsep}}} %
\includegraphics[width=0.235\textwidth, clip]{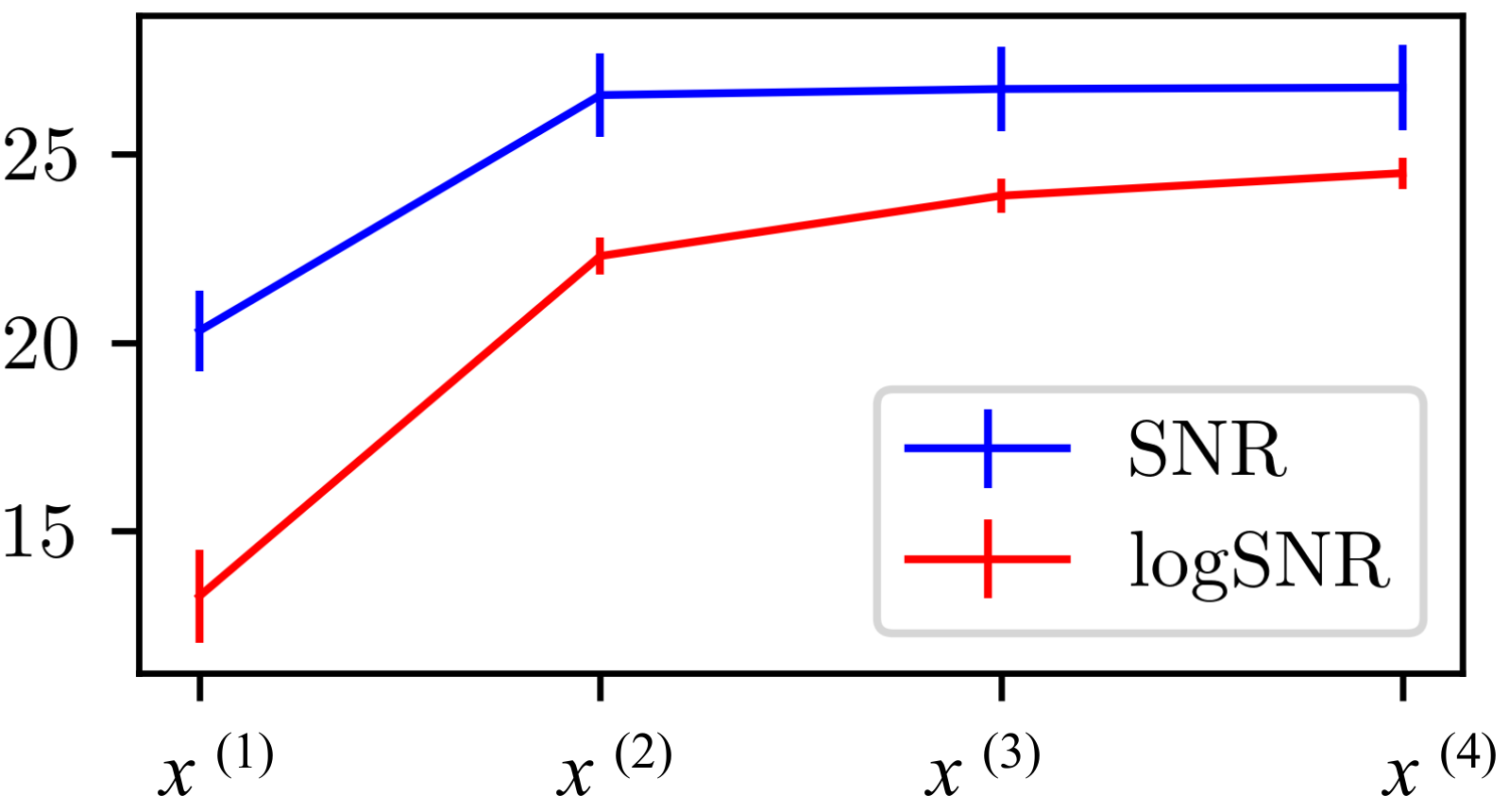} &
\includegraphics[width=0.24\textwidth,  clip]{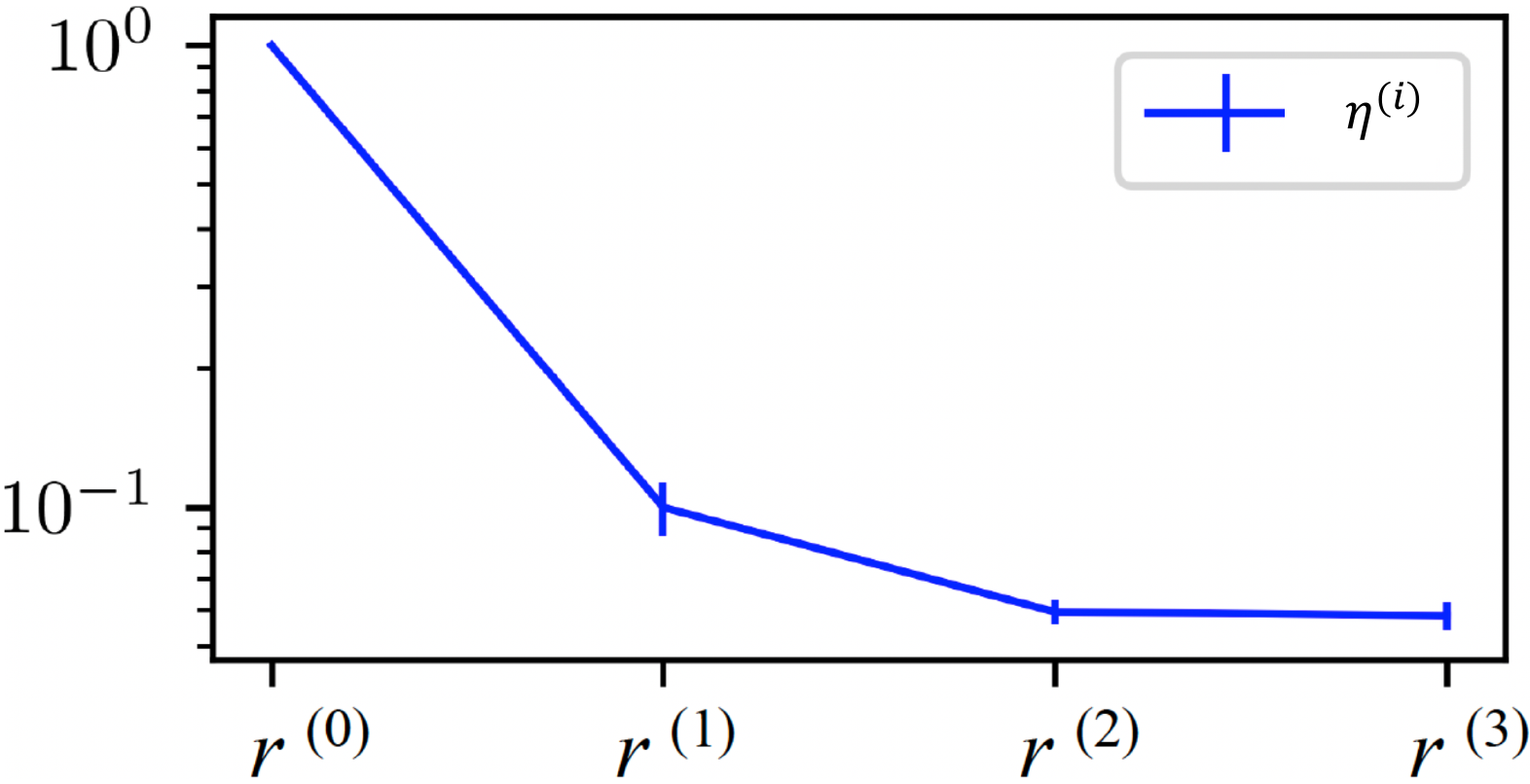} 
\end{tabular}%
\vspace{-0.75em}
\caption{\small Evolution of average reconstruction quality across iterations $1\leq i \leq 4$ over the test dataset, with error bars indicating $95\%$ confidence intervals. Left: SNR and logSNR metrics for $\bm{x}^{(i)}$. Right: Back-projected data residual norm $\eta^{(i)}$.}
\vspace{-0.75em}
\label{fig:metrics_average}%
\end{figure}%

\begin{figure*}[t] \small 
\centering
\begin{tabular}{@{\hspace{0.\tabcolsep}} c @{\hspace{0.05\tabcolsep}} c @{\hspace{0.05\tabcolsep}} c @{\hspace{0.05\tabcolsep}} c @{\hspace{0.05\tabcolsep}} c @{\hspace{0.05\tabcolsep}} c @{\hspace{0.05\tabcolsep}} c @{\hspace{-0.0\tabcolsep}} l @{\hspace{0.\tabcolsep}}}
(a) Ground truth & (b) Back-projected & (c) CLEAN & (d) SARA & (e) AIRI &  (f) $\bm{x}^{(1)}$ & (g) $\bm{x}^{(4)}$ \\
\includegraphics[width=0.135\textwidth]{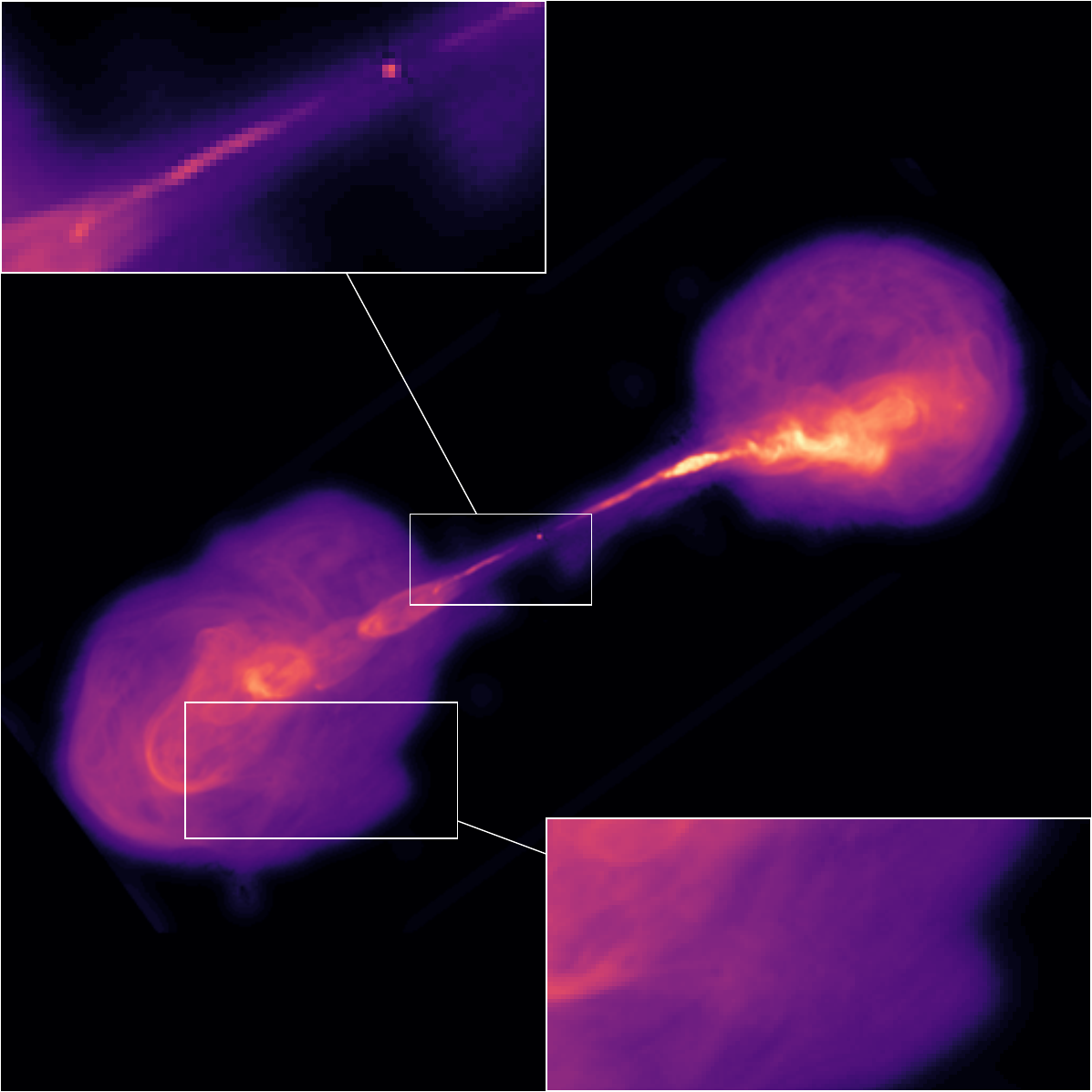} &
\includegraphics[width=0.135\textwidth]{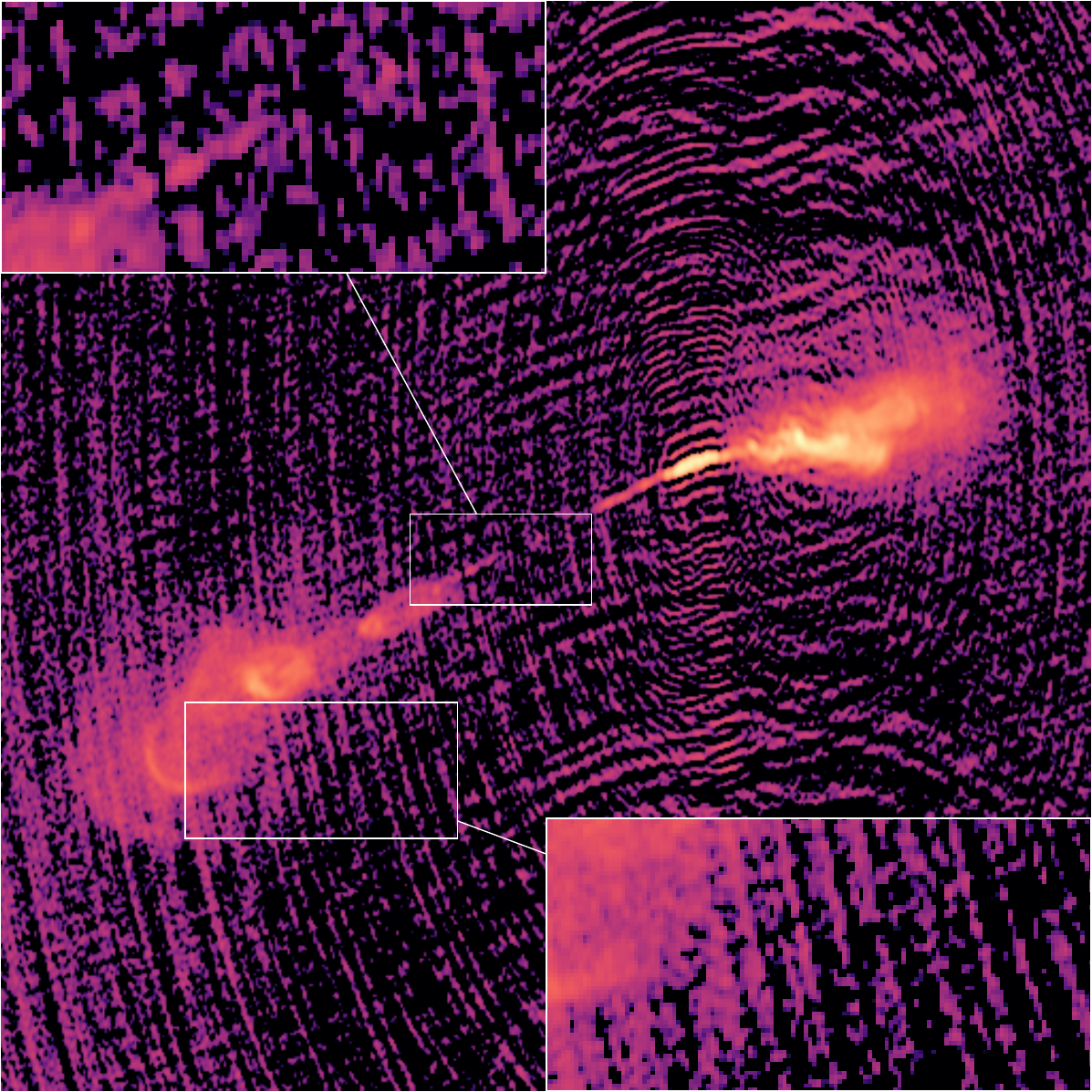} &
\includegraphics[width=0.135\textwidth]{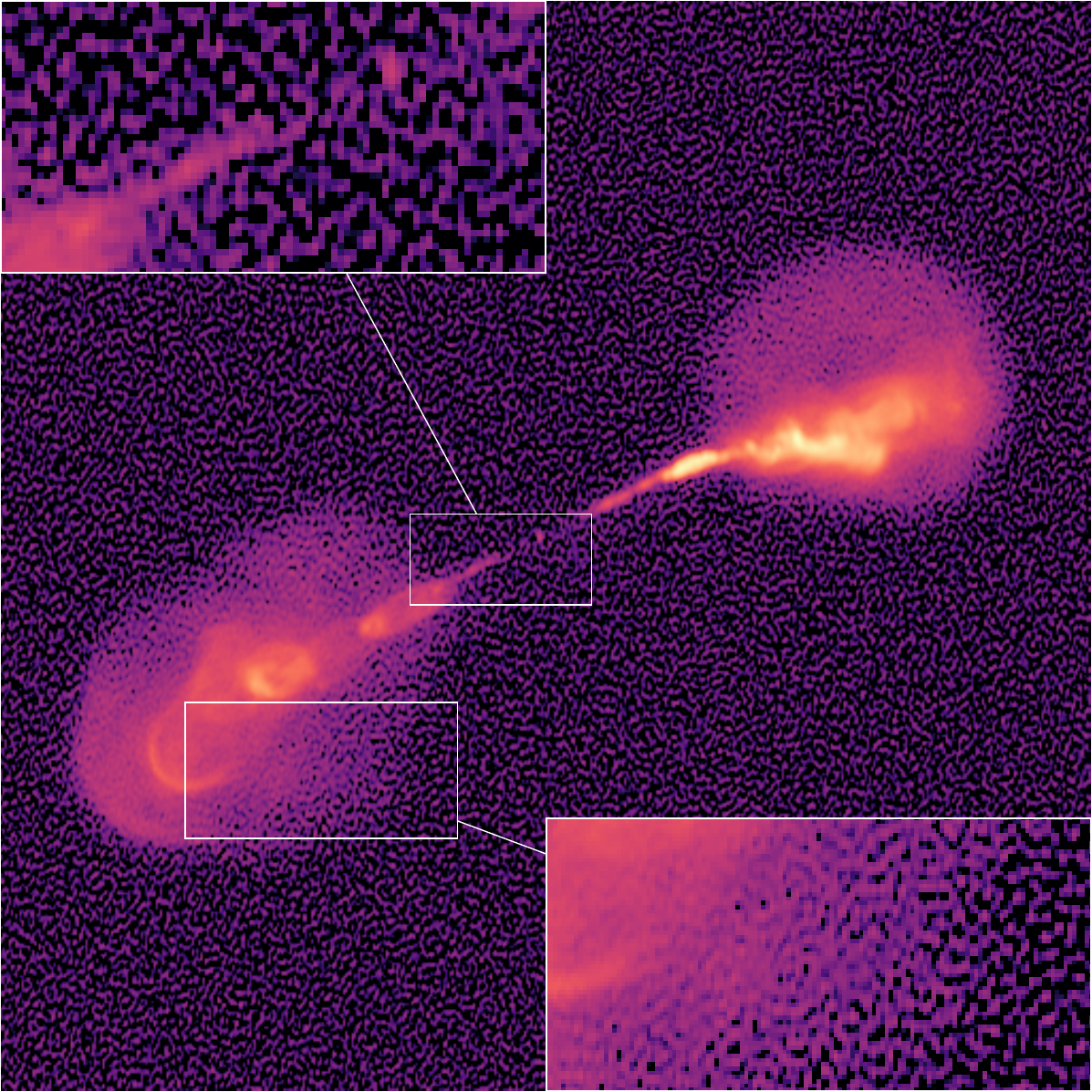} &
\includegraphics[width=0.135\textwidth]{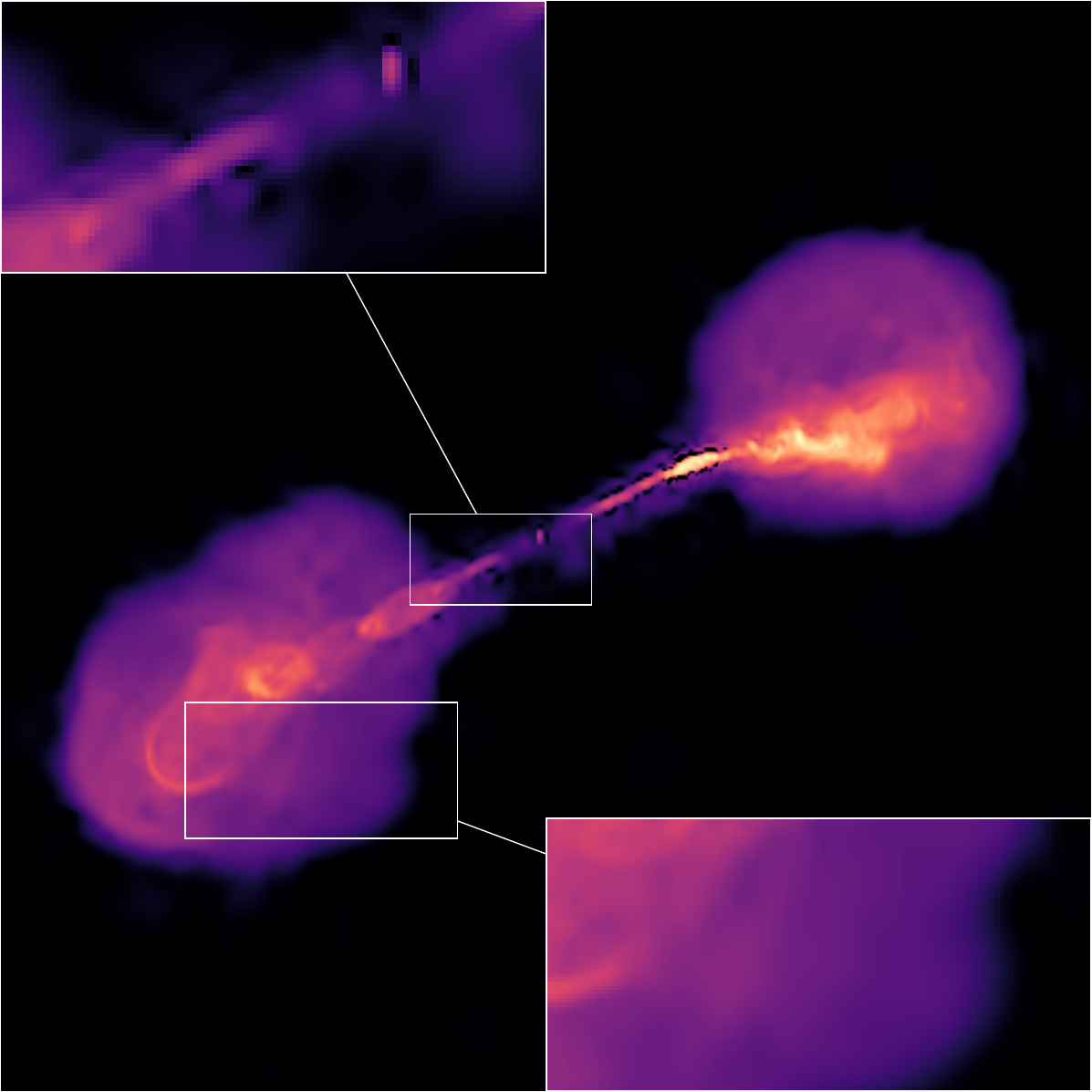} &
\includegraphics[width=0.135\textwidth]{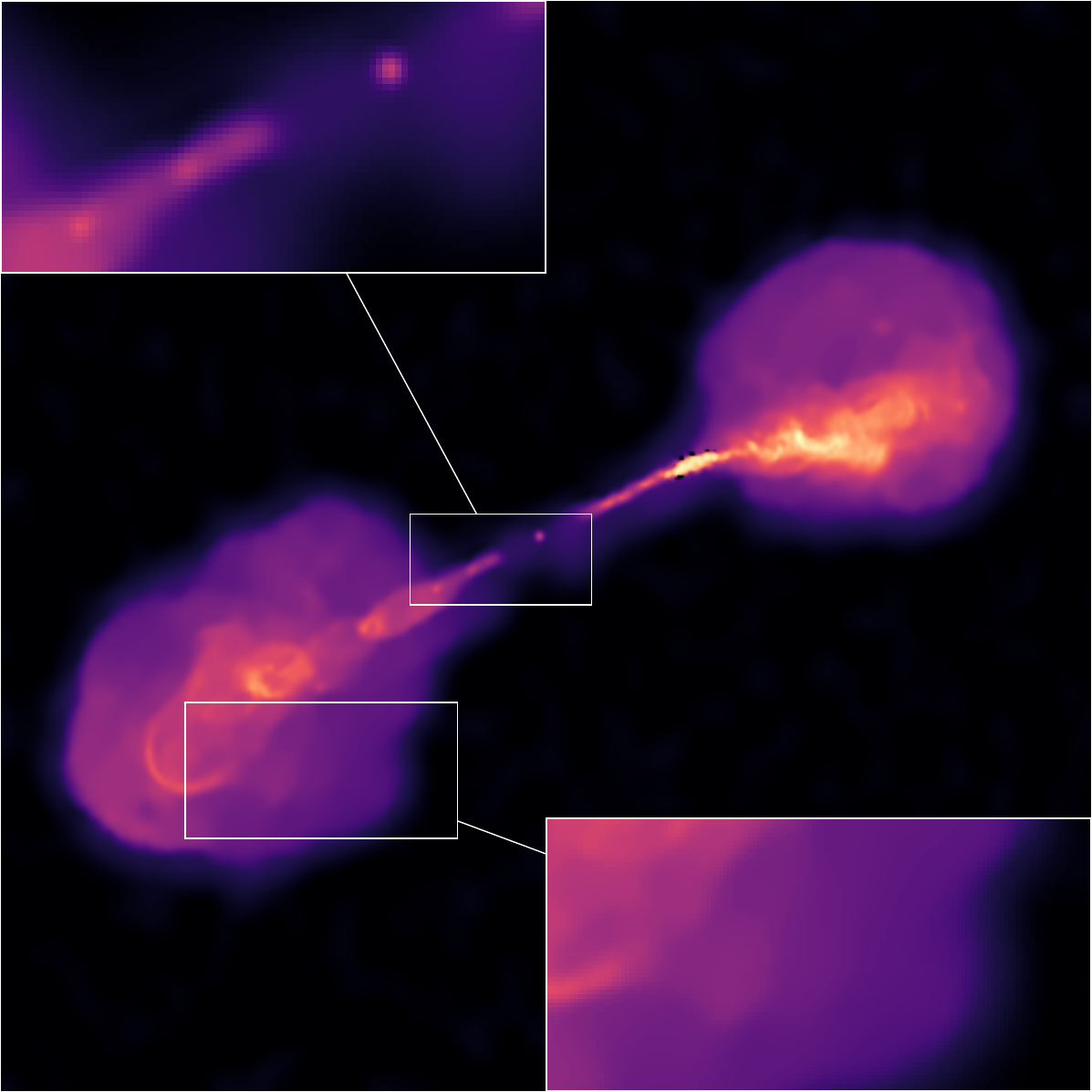} &
\includegraphics[width=0.135\textwidth]{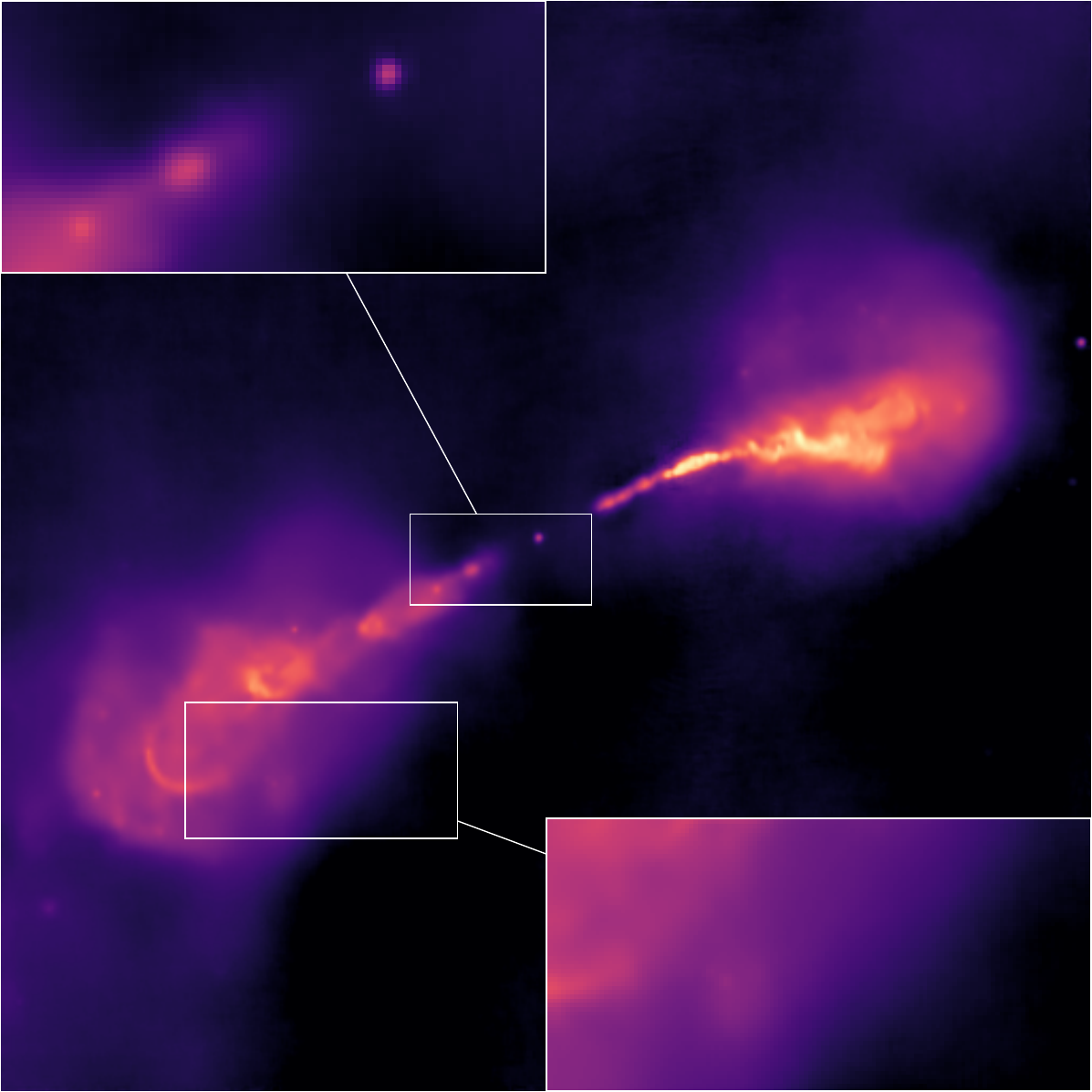} &
\includegraphics[width=0.135\textwidth]{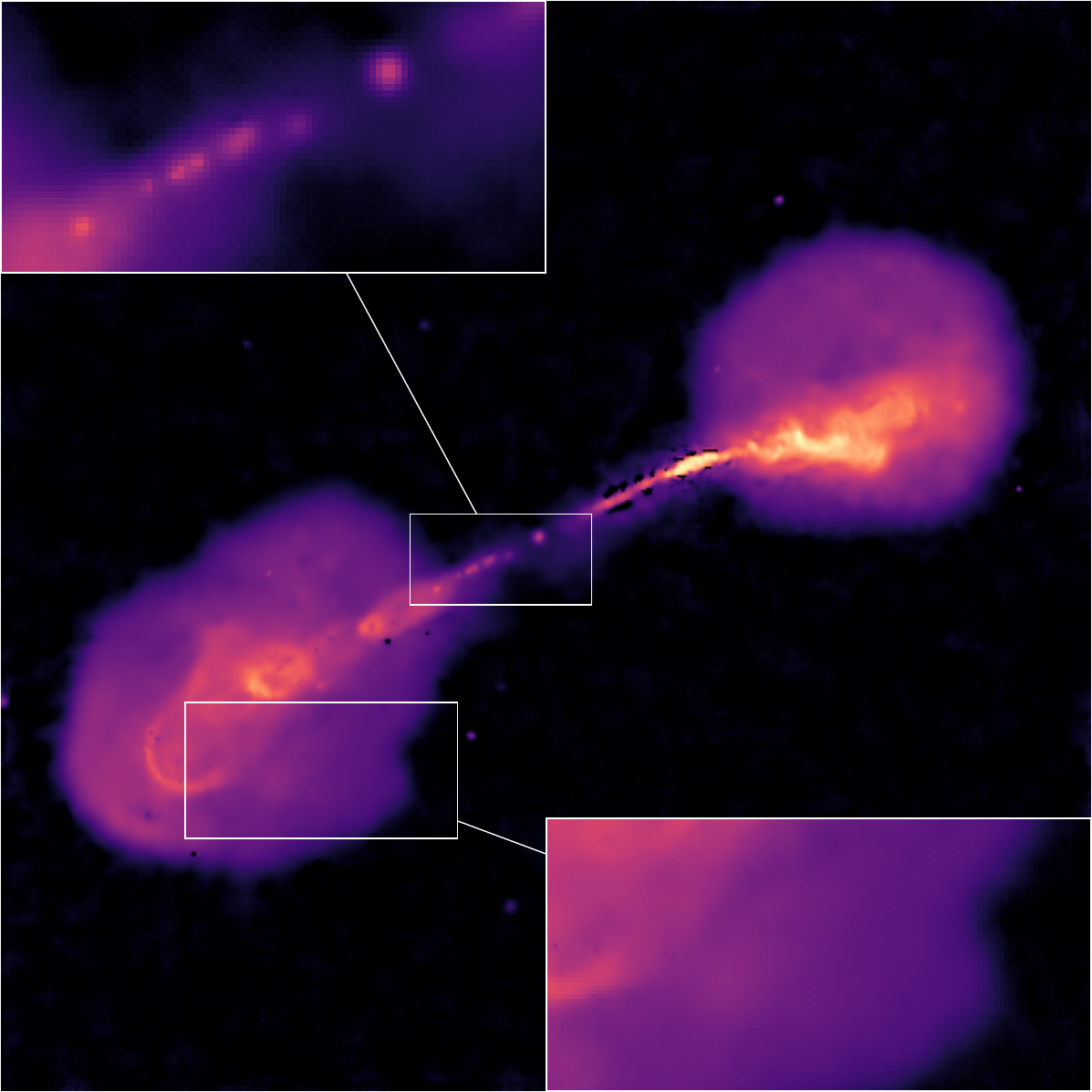} &
\raisebox{-0.5\height}[0pt][0pt]{\includegraphics[width=0.05\textwidth]{images/results/colorbar_vertical.png}} \\
  & $(-11.0, -6.9)$  & $(-2.7, 0.2)$ & $(\mathbf{27.0}, 24.6)$ & $(25.0, 24.7)$ & $(20.8, 11.8)$ & $(26.0, \mathbf{25.1})$\\
 \includegraphics[width=0.135\textwidth]{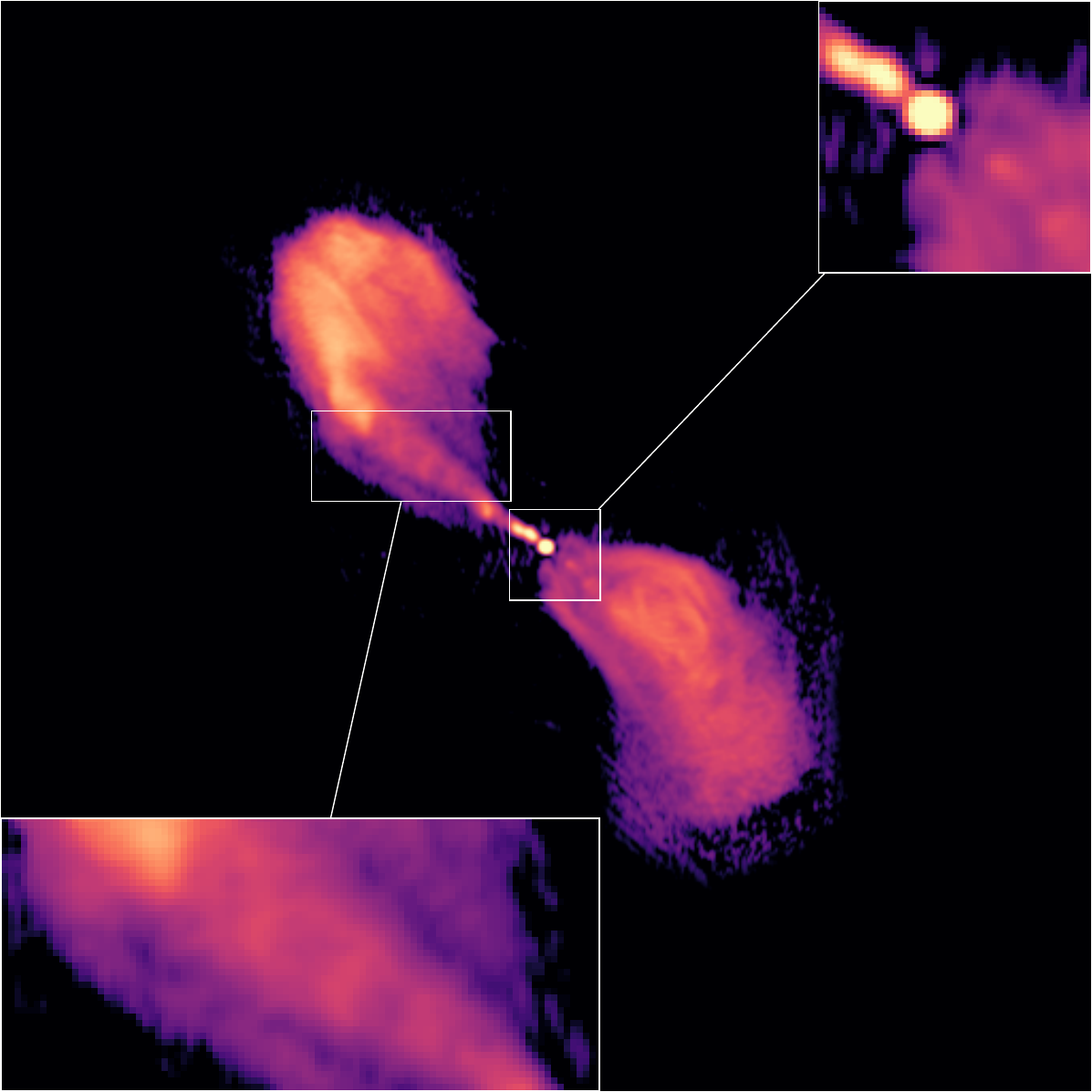} &
\includegraphics[width=0.135\textwidth]{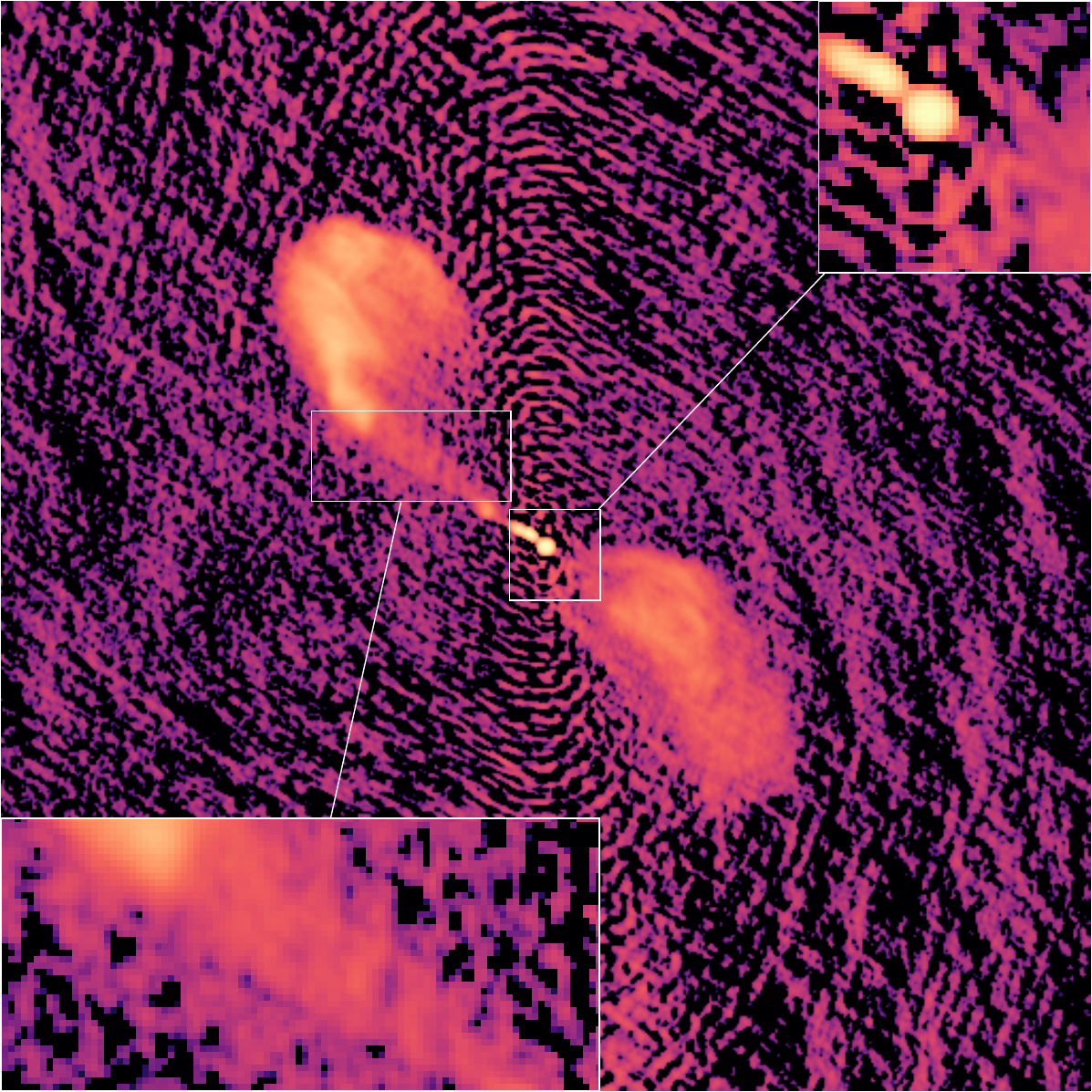} &
\includegraphics[width=0.135\textwidth]{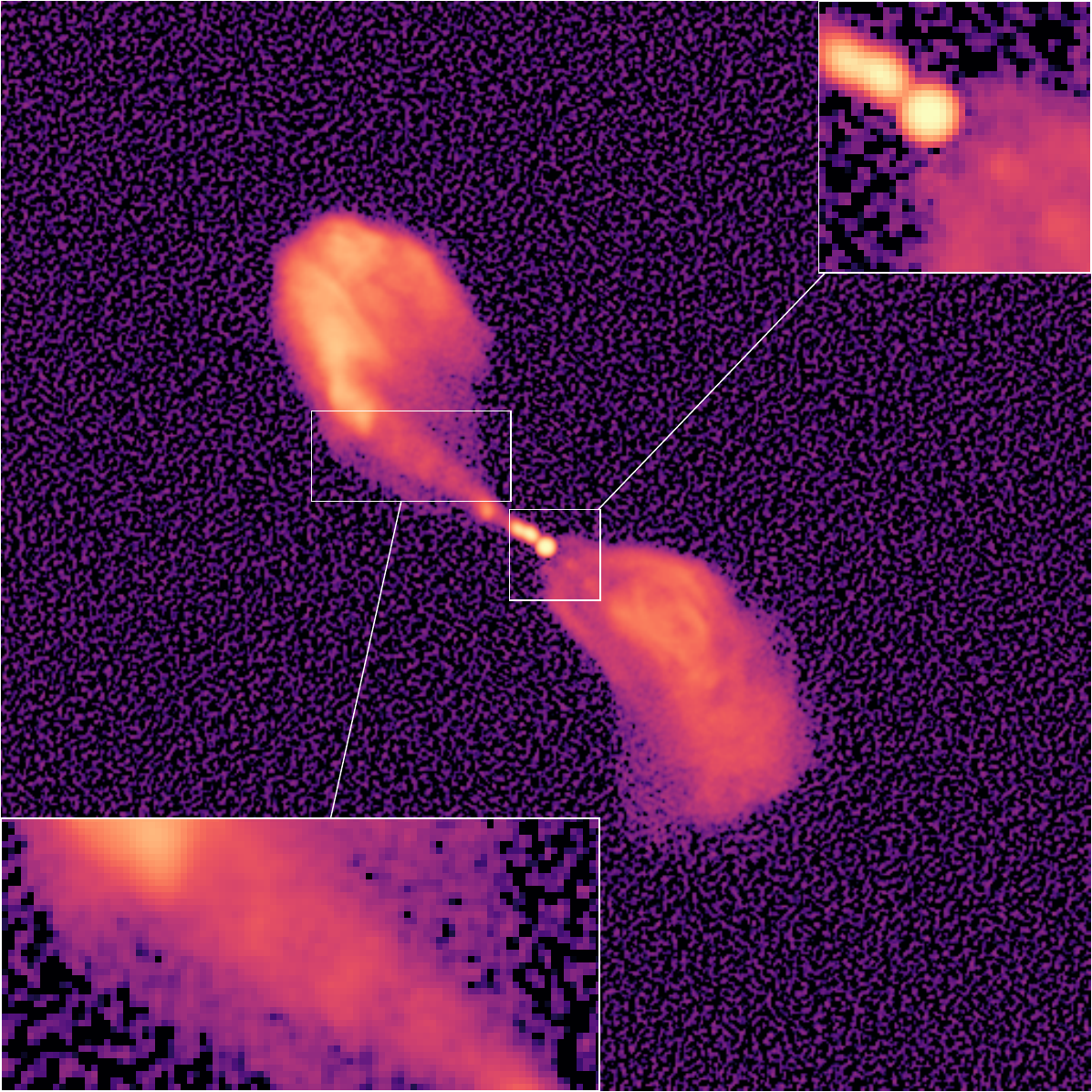} &
\includegraphics[width=0.135\textwidth]{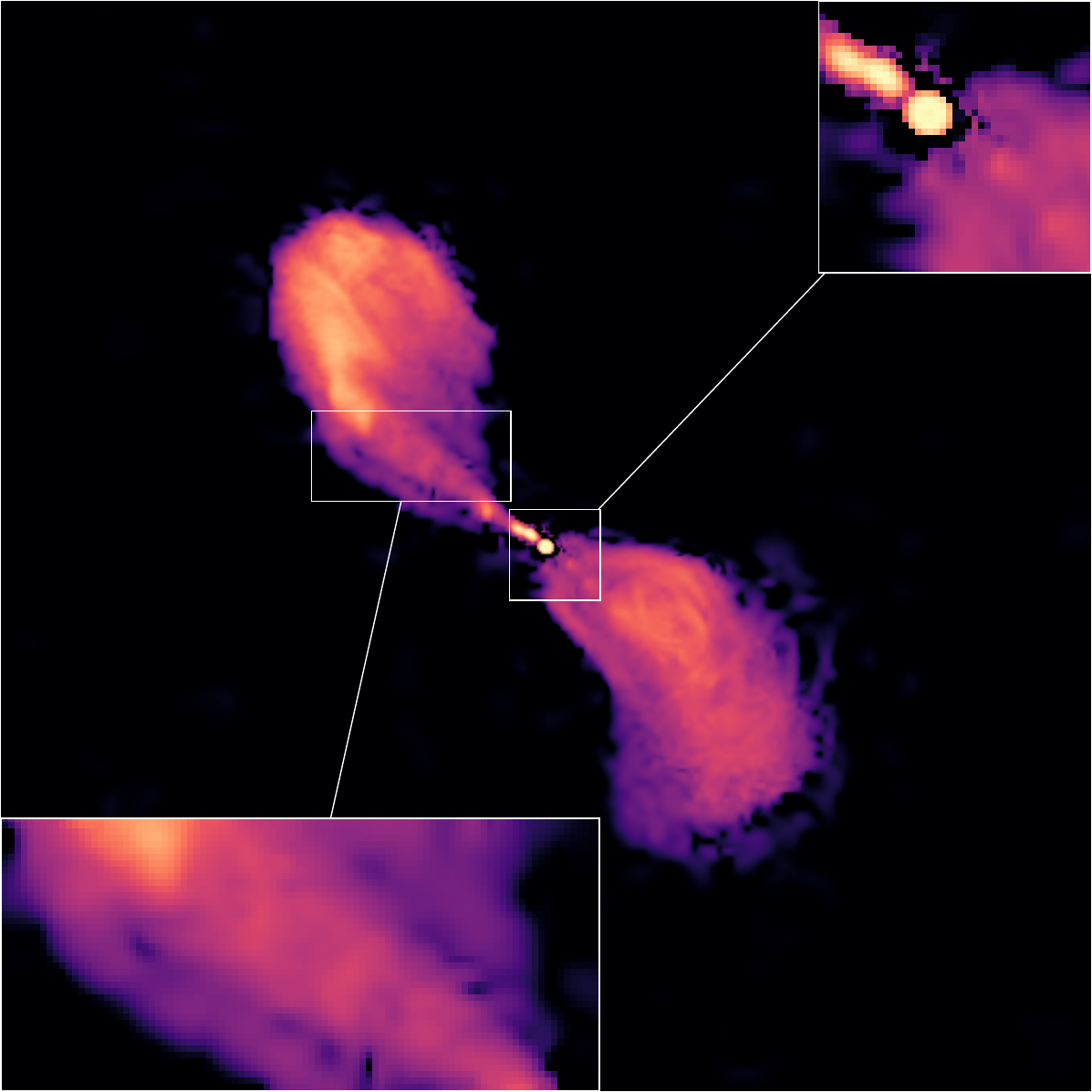} &
\includegraphics[width=0.135\textwidth]{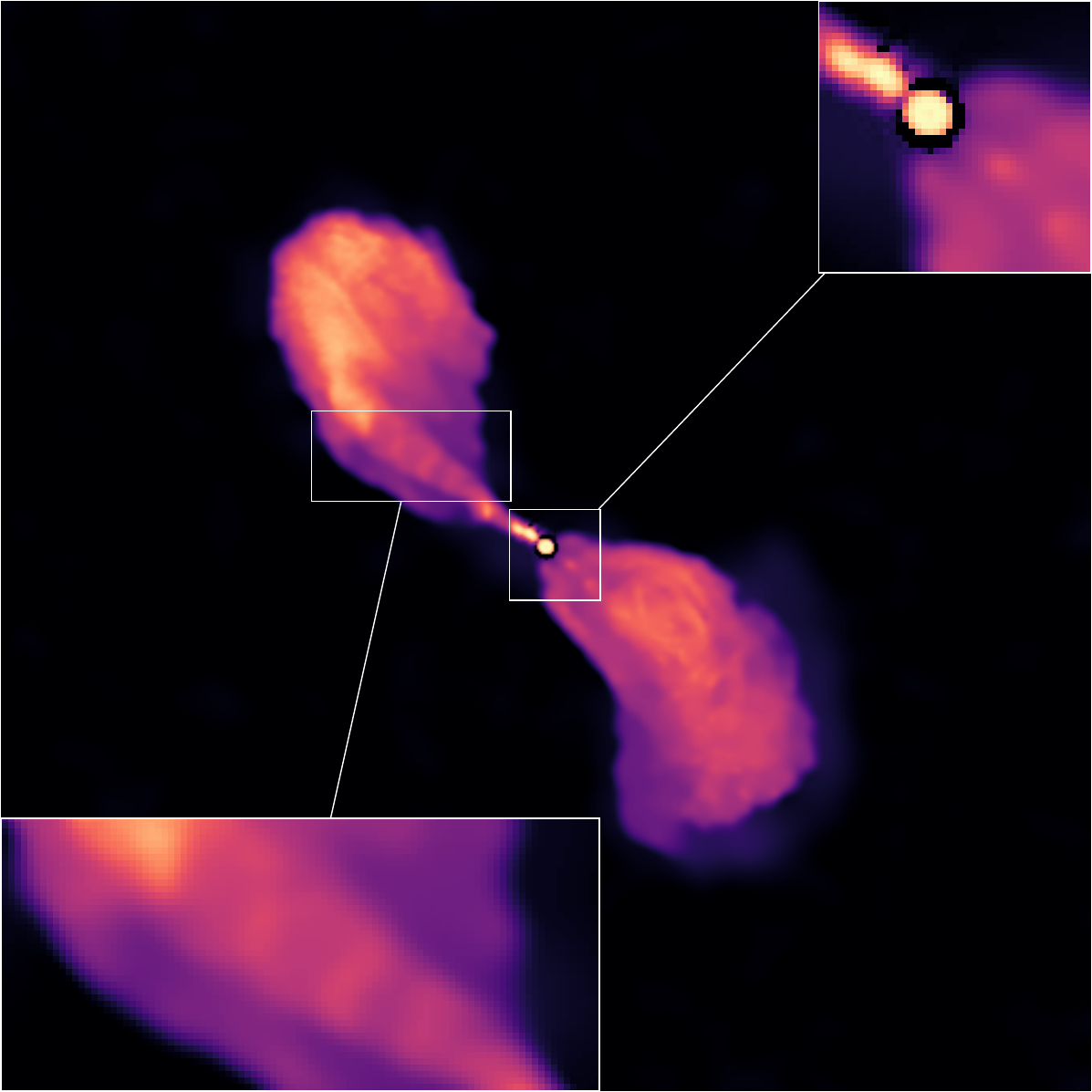} &
\includegraphics[width=0.135\textwidth]{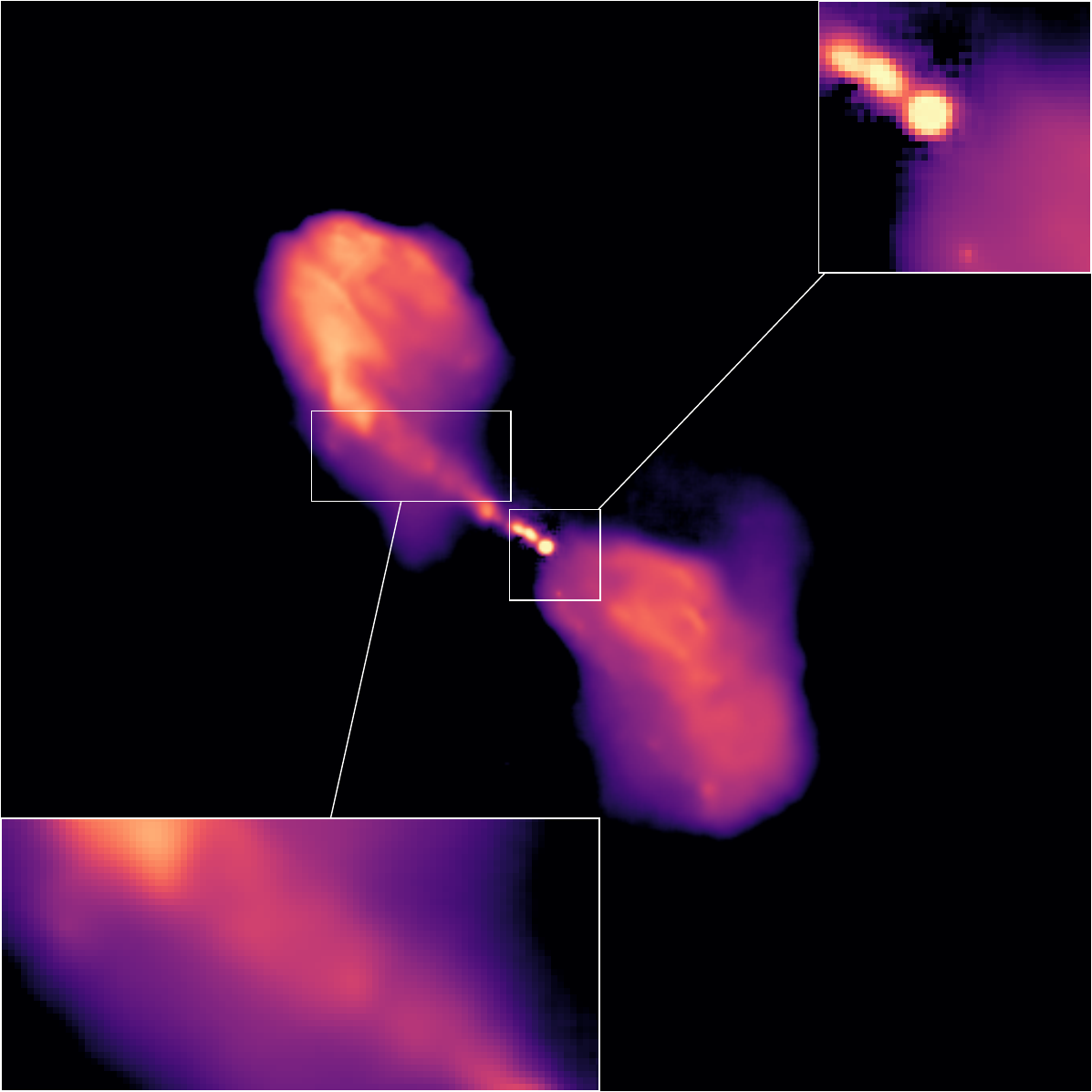} &
\includegraphics[width=0.135\textwidth]{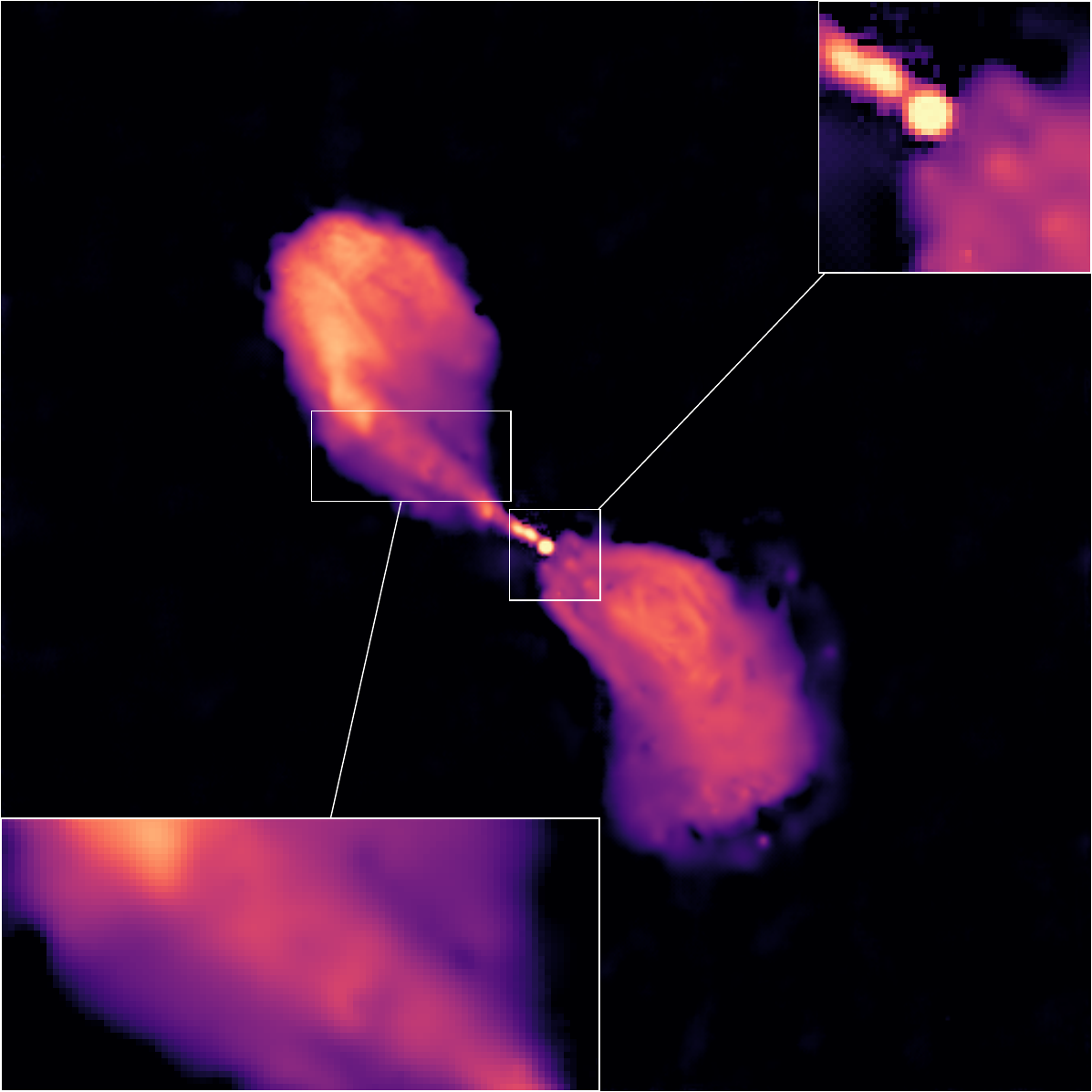} &
 \\
  & $(-13.5, -6.8)$  & $(-4.6, 0.4)$ & $(26.8, \mathbf{24.7})$  & $(25.3, 24.3)$ & $(18.0, 16.9)$ & $(\mathbf{27.7}, 23.8)$ \\
\end{tabular}
\vspace{-1em}
\caption{\small Reconstruction results for the various algorithms considered, for the Hercules A (top) and Centaurus A (bottom) test images, each probed through one of the 5 test MeerKAT sampling patterns. (SNR, logSNR) metrics (in dB) are displayed below each reconstruction.}
\vspace{-0.75em}
\label{fig:comparison_3c353_dt8}
\end{figure*}


\begin{table}[t]
\centering
\small
\begin{tabular}{@{\hspace{0.1\tabcolsep}}l@{\hspace{1.1\tabcolsep}}c@{\hspace{1.1\tabcolsep}}c@{\hspace{1.1\tabcolsep}}c@{\hspace{1.1\tabcolsep}}c@{\hspace{0.1\tabcolsep}}}
\toprule
Algorithm & SNR (dB)  & logSNR (dB) & time (s.) & iteration \#\\
\midrule
CLEAN & $5.3 \pm 0.1$ & $8.2 \pm 0.4$ & $217 \pm 5$ & $8.8\pm 0.2$\\
SARA &  $26.4$$ \pm 0.8$ & $\mathbf{24.9} \pm 0.3$ & $ 11291\pm 563$ & $4825\pm190$ \\
AIRI & $24.9$$ \pm 0.8$ & $\mathbf{25.0} \pm 0.4$ & $4515 \pm 103$ & $6000^*$  \\
\midrule
$\bm{x}^{(1)}$ & $20.3 \pm 1.1$ & $13.3 \pm 1.2$ & $\mathbf{1.0}\pm 0.2$ & $1$\\ 
$\bm{x}^{(4)}$ & $\mathbf{26.8} \pm 1.1$ & $24.5 \pm 0.4$ & $3.9 \pm 0.6$ & $4$\\
\bottomrule
\end{tabular}
\caption{\small Reconstruction SNR, logSNR, times, and number of iterations, for the different imaging algorithms considered. Values reported are averages over the test dataset, with associated 95\% confidence intervals. $^*$AIRI systematically reached the maximum number of iterations.}
\vspace{-1.25em}
\label{tab:comparison_table}
\end{table}

Figure~\ref{fig:step_visualisation} shows the different stages of the proposed reconstruction method, on 3c353 and for the sampling pattern shown in panel (c). Figure~\ref{fig:metrics_average} shows the average SNR and logSNR metrics and residual norms on the 15 inverse problems of the test dataset. The first estimate $\bm{x}^{(1)} = \bm{\mathsf{G}}^{(1)}(\bm{r}^{(0)})$ provides good results at high intensities but is not accurate at faint intensities. This translates into relatively high SNR, but poor logSNR values. The associated residual $\bm{r}^{(1)}$ contains important information and is far from random noise. The output $\bm{x}^{(2)}$ is much more accurate than $\bm{x}^{(1)}$, not only at high intensities (improvement of the SNR), but also at fainter intensities (improvement of the logSNR). 
It is also associated with a residual $\bm{r}^{(2)}$ containing much less information than $\bm{r}^{(1)}$, also evidenced by $\eta^{(2)}<\eta^{(1)}$, supporting the observation that the $\bm{x}^{(2)}$ reconstruction is more faithful to the measurements than $\bm{x}^{(1)}$. While the SNR does not evolve significantly between  $\bm{x}^{(2)}$ and  $\bm{x}^{(3)}$, with also $\eta^{(3)}\simeq\eta^{(2)}$, it appears clearly that $\bm{r}^{(3)}$ contains less residual signal structure than $\bm{r}^{(2)}$, suggesting that $\bm{x}^{(3)}$ provides further improvement at faint intensities. This is evidenced by a noticeable improvement in logSNR. A final logSNR increase is brought in $\bm{x}^{(4)}$.

Table~\ref{tab:comparison_table} shows the average SNR, logSNR, reconstruction times, and number of iterations, as well as associated 95\% confidence intervals over the 15 test inverse problems,  with comparison to the benchmark algorithms. Figure~\ref{fig:comparison_3c353_dt8} provides visual reconstruction results of Hercules A and Centaurus A, for 2 sampling patterns different from the one in Figure~\ref{fig:step_visualisation}. In terms of imaging quality, the reported statistics confirm that the proposed approach is on par with SARA and AIRI, all three methods achieving high resolution and high dynamic range. The end-to-end approach falls short from delivering a similar resolution and dynamic range, but is still significantly better than CLEAN. In terms of reconstruction times, the proposed approach is two orders of magnitude faster than CLEAN, and three orders of magnitude faster than AIRI, itself more than twice as fast as SARA. The computation time of each method crudely evolves proportionally to the number of iterations. More precisely, the proposed method, the end-to-end approach, and AIRI, share a very similar computation time per iteration, due to a similar iteration structure consisting in computing a residual followed by application of a DNN. SARA's iteration cost is heavier due to costly proximal operations associated with its regularisation approach. CLEAN's iteration cost is also heavier, mainly due to the fact it does not keep the measurement operator in memory.

We finally note that preliminary results not reported here suggest that the proposed hybrid DNN architecture used outperforms architectures such as UNet and WDSR, both in the end-to-end and residual network series approaches.

\section{Conclusion}
\label{sec:conclusion}

In this paper, we proposed a novel residual DNN series approach for solving large-scale high-dynamic range computational imaging problems, circumventing scalability challenges encountered by both unfolded end-to-end architectures and PnP approaches. Each network of the series, trained to transform a back-projected data residual into an image residual, improves the dynamic range from the previous iteration, with the sum of the output residual images across iterations representing the final reconstruction.

Simulation results for RI imaging show that a series of only a few terms achieves similar reconstruction quality to the state-of-the-art optimisation algorithm SARA and PnP method AIRI, whilst being orders of magnitude faster.

Future work should investigate the robustness of the approach to the wide variations of measurements settings characterising RI observation, and its practical effectiveness on real large-scale high-dynamic range data. The suggested superiority of the DNN architecture proposed over state-of-the-art architectures should also be confirmed. The application of the proposed methodology to other computational imaging modalities is naturally an open question.

\bibliographystyle{IEEEbib}
\bibliography{main}

\begin{thebibliography}{10}

\bibitem{wiaux2009compressed}
Yves Wiaux, Laurent Jacques, Gilles Puy, Anna~MM Scaife, and Pierre
  Vandergheynst,
\newblock ``Compressed sensing imaging techniques for radio interferometry,''
\newblock {\em Monthly Notices of the Royal Astronomical Society}, vol. 395,
  no. 3, pp. 1733--1742, 2009.

\bibitem{terris2022image}
Matthieu Terris, Arwa Dabbech, Chao Tang, and Yves Wiaux,
\newblock ``Image reconstruction algorithms in radio interferometry: From
  handcrafted to learned regularization denoisers,''
\newblock {\em Monthly Notices of the Royal Astronomical Society}, 2022.

\bibitem{ahmad2020plug}
Rizwan Ahmad, Charles~A Bouman, Gregery~T Buzzard, Stanley Chan, Sizhuo Liu,
  Edward~T Reehorst, and Philip Schniter,
\newblock ``Plug-and-play methods for magnetic resonance imaging: Using
  denoisers for image recovery,''
\newblock {\em IEEE signal processing magazine}, vol. 37, no. 1, pp. 105--116,
  2020.

\bibitem{muckley2021results}
Matthew~J Muckley, Bruno Riemenschneider, Alireza Radmanesh, Sunwoo Kim, Geunu
  Jeong, Jingyu Ko, Yohan Jun, Hyungseob Shin, Dosik Hwang, Mahmoud Mostapha,
  et~al.,
\newblock ``{Results of the 2020 fastMRI challenge for machine learning MR
  image reconstruction},''
\newblock {\em IEEE transactions on medical imaging}, vol. 40, no. 9, pp.
  2306--2317, 2021.

\bibitem{liang2021swinir}
Jingyun Liang, Jiezhang Cao, Guolei Sun, Kai Zhang, Luc Van~Gool, and Radu
  Timofte,
\newblock ``Swinir: Image restoration using swin transformer,''
\newblock in {\em IEEE/CVF International Conference on Computer Vision}, 2021,
  pp. 1833--1844.

\bibitem{gheller2021convolutional}
Claudio Gheller and Franco Vazza,
\newblock ``Convolutional deep denoising autoencoders for radio astronomical
  images,''
\newblock {\em Monthly Notices of the Royal Astronomical Society}, 2021.

\bibitem{connor2022deep}
Liam Connor, Katherine~L Bouman, Vikram Ravi, and Gregg Hallinan,
\newblock ``Deep radio-interferometric imaging with polish: Dsa-2000 and weak
  lensing,''
\newblock {\em Monthly Notices of the Royal Astronomical Society}, vol. 514,
  no. 2, pp. 2614--2626, 2022.

\bibitem{adler2018learned}
Jonas Adler and Ozan {\"O}ktem,
\newblock ``Learned primal-dual reconstruction,''
\newblock {\em IEEE Transactions on Medical Imaging}, vol. 37, no. 6, pp.
  1322--1332, 2018.

\bibitem{banert2020data}
Sebastian Banert, Axel Ringh, Jonas Adler, Johan Karlsson, and Ozan Oktem,
\newblock ``Data-driven nonsmooth optimization,''
\newblock {\em SIAM Journal on Optimization}, vol. 30, no. 1, pp. 102--131,
  2020.

\bibitem{venkatakrishnan2013plug}
Singanallur~V Venkatakrishnan, Charles~A Bouman, and Brendt Wohlberg,
\newblock ``Plug-and-play priors for model based reconstruction,''
\newblock in {\em 2013 IEEE Global Conference on Signal and Information
  Processing}. IEEE, 2013, pp. 945--948.

\bibitem{zhang2021plug}
Kai Zhang, Yawei Li, Wangmeng Zuo, Lei Zhang, Luc Van~Gool, and Radu Timofte,
\newblock ``Plug-and-play image restoration with deep denoiser prior,''
\newblock {\em IEEE Transactions on Pattern Analysis and Machine Intelligence},
  2021.

\bibitem{pesquet2021learning}
Jean-Christophe Pesquet, Audrey Repetti, Matthieu Terris, and Yves Wiaux,
\newblock ``Learning maximally monotone operators for image recovery,''
\newblock {\em SIAM Journal on Imaging Sciences}, vol. 14, no. 3, pp.
  1206--1237, 2021.

\bibitem{dabbech2022first}
Arwa Dabbech, Matthieu Terris, Adrian Jackson, Mpati Ramatsoku, Oleg~M Smirnov,
  and Yves Wiaux,
\newblock ``{First AI for deep super-resolution wide-field imaging in radio
  astronomy: unveiling structure in ESO 137-006},''
\newblock {\em Accepted for publication in The Astrophysical Journal Letters},
  2022.

\bibitem{bauschke2017convex}
Heinz~H Bauschke and Patrick~L Combettes,
\newblock {\em Convex analysis and monotone operator theory in Hilbert spaces},
\newblock Springer, 2017.

\bibitem{gilton2019neumann}
Davis Gilton, Greg Ongie, and Rebecca Willett,
\newblock ``Neumann networks for linear inverse problems in imaging,''
\newblock {\em IEEE Transactions on Computational Imaging}, vol. 6, pp.
  328--343, 2019.

\bibitem{yu2018wide}
Jiahui Yu, Yuchen Fan, Jianchao Yang, Ning Xu, Zhaowen Wang, Xinchao Wang, and
  Thomas Huang,
\newblock ``Wide activation for efficient and accurate image
  super-resolution,''
\newblock {\em arXiv preprint arXiv:1808.08718}, 2018.

\bibitem{onose2016scalable}
Alexandru Onose, Rafael~E Carrillo, Audrey Repetti, Jason~D McEwen,
  Jean-Philippe Thiran, Jean-Christophe Pesquet, and Yves Wiaux,
\newblock ``Scalable splitting algorithms for big-data interferometric imaging
  in the ska era,''
\newblock {\em Monthly Notices of the Royal Astronomical Society}, vol. 462,
  no. 4, pp. 4314--4335, 2016.

\bibitem{thouvenin2022parallel}
Pierre-Antoine Thouvenin, Abdullah Abdulaziz, Arwa Dabbech, Audrey Repetti, and
  Yves Wiaux,
\newblock ``Parallel faceted imaging in radio interferometry via proximal
  splitting (faceted hypersara): I. algorithm and simulations,''
\newblock {\em Monthly Notices of the Royal Astronomical Society}, 2022.

\bibitem{offringa2014wsclean}
AR~Offringa, Benjamin McKinley, Natasha Hurley-Walker, FH~Briggs, RB~Wayth,
  DL~Kaplan, ME~Bell, Lu~Feng, AR~Neben, JD~Hughes, et~al.,
\newblock ``{WSCLEAN: an implementation of a fast, generic wide-field imager
  for radio astronomy},''
\newblock {\em Monthly Notices of the Royal Astronomical Society}, vol. 444,
  no. 1, pp. 606--619, 2014.

\end{thebibliography}

\end{document}